\documentclass[journal]{IEEEtran}
\usepackage{geometry}
\pdfoutput=1
\ifCLASSINFOpdf

\else

\fi
\usepackage[cmex10]{amsmath}
\usepackage{amssymb}
\usepackage{graphicx}
\usepackage{array,color}
\usepackage{amsmath}
\usepackage{stfloats}
\usepackage{graphicx}
\usepackage{subfigure}
\usepackage{tabularx}
\usepackage{epsfig,epsf,color,balance}
\usepackage{algorithm}
\usepackage{algorithmicx}
\usepackage{algpseudocode}
\usepackage{algpseudocode}\floatname{algorithm}{Algorithm}   
\usepackage{bm}
\usepackage{textcomp}
\usepackage{amsmath}
\usepackage[justification=centering]{caption}
\usepackage{cite}
\usepackage[citecolor=green]{hyperref}

\begin{document}

\title{Reconfigurable Intelligent Surface-Aided Dual-Function Radar and Communication Systems With  MU-MIMO Communication}

\author{\IEEEauthorblockN{Yasheng Jin, Hong Ren, \IEEEmembership{Member, IEEE}, Cunhua Pan, \IEEEmembership{Senior, IEEE}, Zhiyuan Yu, Ruisong Weng, Boshi Wang,  Gui Zhou, Yongchao He and Maged Elkashlan, \IEEEmembership{Senior, IEEE}}\\

\thanks{Y. Jin,  H. Ren, C. Pan, Z. Yu, R. Weng, B. Wang and Y. He are with National Mobile Communications Research Laboratory, Southeast University, Nanjing, China. (e-mail:\{yashengjin, zyyu, ruisong\_weng, boshiwang, hren, cpan,  heyongchao\}@seu.edu.cn). G. Zhou is with the Institute for Digital Communications, Friedrich-Alexander-University Erlangen-Nürnberg (FAU), 91054 Erlangen, Germany (e-mail: gui.zhou@fau.de). M. Elkashlan is with the School of Electronic Engineering and Computer Science, Queen Mary University of London, E1 4NS London, U.K. (e-mail: maged.elkashlan@qmul.ac.uk).
}  
\thanks{Corresponding author: Hong Ren and Cunhua Pan. }
}

%\thanks{C. Pan, H. Ren, M. Elkashlan and A. Nallanathan are with the School of Electronic Engineering and Computer Science at  Queen Mary University of London, London E1 4NS, U.K. (e-mail:\{c.pan, h.ren, maged.elkashlan, a.nallanathan\}@qmul.ac.uk). K. Wang is with Department of Computer and Information Sciences, Northumbria University, UK. (e-mail: kezhi.wang@northumbria.ac.uk).   W. Xu is with National Mobile Communications Research Laboratory, Southeast University, Nanjing 210096, China. (e-mail: wxu@seu.edu.cn).  L. Hanzo is with the School of Electronics and Computer Science, University of Southampton, Southampton, SO17 1BJ, U.K. (e-mail: lh@ecs.soton.ac.uk). }	
	
%	National Mobile Communications Research Laboratory Southeast University Nanjing, China} \\
%		Email: \{yashengjin, zyyu, ruisong\_weng, boshiwang, hren, cpan\}@seu.edu.cn}
%	Z. Yu, H. Ren, C. Pan and B. Wang are with National Mobile Communications Research Laboratory, Southeast University, Nanjing, China. (e-mail:\{zyyu, hren, cpan, boshiwang\}  @seu.edu.cn).

\maketitle
\vspace{-1.9cm}
\begin{abstract}
In this paper, we investigate an reconfigurable intelligent surface (RIS)-aided integrated sensing and communication (ISAC) system. Our objective is to maximize the achievable sum rate of the multi-antenna communication users through the joint active and passive beamforming. {Specifically}, the weighted minimum mean-square error (WMMSE) method is { first} used to reformulate the original problem into an equivalent one. Then, we utilize an alternating optimization (AO) { algorithm} to decouple the optimization variables and decompose this challenging problem into two subproblems. Given reflecting coefficients, a penalty-based algorithm is utilized to  deal with  
%the transmit power  and 
{ the non-convex radar signal-to-noise ratio (SNR) constraints}. For the  given beamforming matrix of the BS, we apply majorization-minimization (MM) to  transform the  problem into a quadratic constraint quadratic programming (QCQP) problem, which is ultimately solved using a semidefinite relaxation (SDR)-based algorithm. Simulation results illustrate the advantage of deploying RIS in the considered multi-user MIMO (MU-MIMO) ISAC systems.

\end{abstract}


\IEEEpeerreviewmaketitle
%\index{}
%\begin{multicols}{2}
\section{Introduction}\label{intro}

The evolution of { higher frequency} bands in wireless communication, such as terahertz and visible light, is progressively converging with the established frequency bands used for traditional sensing applications \cite{7409935}.
%The development of higher frequency bands for wireless communications, such as terahertz and visible light, will increasingly overlap with traditional sensing bands\cite{7409935}. 
{ Integrated sensing and communication} (ISAC) offers the capability to seamlessly integrate communication and sensing within the same spectrum, thereby mitigating interference and enhancing spectrum utilization, which is the preferred path for technology and industry\cite{ma2020joint}. The communication radar coexistence (CRC)  and the  dual-function radar-communication (DFRC) systems are two different directions of ISAC research. The goal of CRC is to achieve the coexistence of communication and sensing functions on the same frequency bandwidth while the hardware systems of communication and radar are entirely independent\cite{zheng2019radar,7470514}. Compared with CRC, DFRC systems can implement communication and radar sensing functions on the same system \cite{liu2020jointApplications}, significantly improving spectrum utilization and hardware efficiency while also reducing system power consumption\cite{liu2018mimo,hassanien2015dual}.

One of the keys and challenges to achieving an efficient DFRC system is proper waveform design. Generally, existing literature can be classified into three categories\cite{9924202}: communication-centric waveform design (CCWD)\cite{sturm2011waveform,liyanaarachchi2021optimized,zeng2020joint,wu2021otfs}, sensing-centric waveform design (SCWD)\cite{xie2021waveform,nowak2016co,wang2022vehicle}, and joint waveform optimization and design (JWOD)\cite{liu2020joint}. The core objective of CCWD is to adapt conventional communication waveforms, such as orthogonal frequency division multiplexing (OFDM) waveform and orthogonal time frequency space (OTFS) waveform, to have a particular capability of  sensing targets. 
%The primary purpose of CCWD is to modify existing communication waveforms, such as orthogonal frequency division multiplexing (OFDM) waveform and orthogonal time frequency space (OTFS) waveform, to have a particular ability to sensing targets. 
Specifically, these modified waveforms can obtain the desired sensing information from the echo. However, due to the randomness brought about by communication data, it may significantly impact sensing performance.
Unlike CCWD, SCWD aims to design the sensing waveform so that the modified waveform has communication capabilities. Generally speaking, we can embed communication information into the sensing waveform, for example: embedding communication signals into the spatial domain to realize SCWD\cite{yang2020dual}.
Although embedding communication symbols into radar waveforms is straightforward and convenient, the low data transmission rate problem brought by this method cannot be ignored, making it only applicable to specific scenarios. Besides, the lack for practical principles for demodulation is also a challenging problem for embedded communication information.
%Although almost any radar waveform can be used to embed the communication symbols in different resources/blocks, such as pulse, frequency, and phase, the performance degradation of the radar sensing caused by the communication information seems inevitable to some extent. 
%Besides, due to the embedding information, the transceiver hardware architecture will inevitably become more complex to code/decode communication symbols. It is also worth considering the ways to reduce the hardware cost.
In contrast to CCWD and SCWD, JWOD does not modify the existing communication or radar waveforms but redesigns waveforms based on actual scenarios, making this method more flexible. Generally speaking, new waveforms have more spatial degrees of freedom (DoF), achieving a balance of performance between communication and perception.
%
%without being constrained to existing waveforms, JWOD exhibits more DoFs and flexibility to balance the requirements of sensing and communication, potentially providing a better tradeoff between the two functionalities, even improving the performance simultaneously. 
The work of \cite{10005137} proposed a beamforming design for multi-user DFRC systems, and it has shown that appropriately designed radar waveforms can increase the DoF used for target detection.

The performance of the DFRC system may be significantly deteriorated by unfavorable propagation environments with signal blockages, especially for target sensing. Fortunately, reconfigurable intelligent surface (RIS) can address this issue by manipulating the wireless propagation environment with low power consumption and hardware cost\cite{ wu2019towards}. RIS is usually composed of many passive, low-cost elements, each of which can independently adjust the phase of the  incident signal\cite{cui2014coding,yu2011light}. The great advantage of the RIS has been widely verified in many fields, such as extending the coverage area and enhancing the reliability\cite{9847080,pan2020multicell}.

The application of RIS in other fields of communication also inspires researchers to combine it with DFRC system. In the early stages, the RIS is exclusively used to enhance communication functionality, with direct transceiver-target links employed for sensing tasks\cite{9838546,9416177}.
%The application of RIS in the field of communication also inspires researchers to combine it with DFRC system. From the perspective of the role that RIS plays in DFRC, in the simplest case, RIS may only be used to assist in communication. For example, in DFRC systems with a strong   line-of-sight (LoS) link between the BS and the target, RIS is generally only used to assist communication\cite{liu2022deep,9416177}. 
The work of \cite{9838546} studied an RIS-assisted ISAC system with an ISAC BS, multiple targets and  communication users (CUs). RIS is deployed in the system to provide CUs with a feasible BS-RIS-CU link to  improve the signal strength in the communication coverage area. The authors of \cite{9416177} studied an RIS-assisted ISAC system where dual-function BS simultaneously serves multiple CUs and multiple sensing targets, while RIS is only used to assist communication to minimize multi-user interference (MUI). 

However, in urban environment, surrounding buildings are likely to block the BS-target path. To further leverage the benefits of RIS, especially in augmenting radar sensing performance, RIS is also used to establish  a virtual link between the BS and the target.
%However, in an urban environment, surrounding buildings are likely to hinder the BS-target path. In this case, properly deploying RIS can provide the target with an additional transmission link to support radar sensing performance.
%In DFRC system, it can change the direction of the reflected signal by adjusting the phase shift of the elements to help the system provide a non-LoS (NLoS) link for the blocked sense target, and it can also improve the communication quality-of-service (QoS).
%In recent studies, RIS  is more used in ISAC systems to assist both communication and target sensing simultaneously. 
%In \cite{jiang2021intelligent}, both target sensing and single antenna communication were enhanced by deploying RIS near the BS. A more general multi-target sensing task in RIS-aided ISAC systems was considered in \cite{song2022joint}. Specifically, the minimum beampattern gain in the perceptual target direction is maximized while serving a single antenna communication user (CU) simultaneously.
%In the following, the work of  \cite{luo2023ris} and \cite{zhang2022joint} studied more realistic application scenarios with  multiple targets and multiple CUs. The authors of \cite{luo2023ris} aimed to maximize the weighted radar sum signal-to-noise ratio (SNR)
%with RIS-aided multiple potential targets and multiple users. 
%In \cite{zhang2022joint}, a unified multiple target and multiple users ISAC systems was studied. Furthmore,  a mutual information (MI) based optimization method is proposed, obviating the need for individual channel information for target detection.
The work of \cite{jiang2021intelligent} deployed RIS near the BS, and the BS serves  one single-antenna CU while detecting a target, where the signal-to-noise ratio (SNR) of radar is maximized while satisfying the SNR constraints of CUs. 
%To mitigate the serious problem of channel fading, we propose deploying an IRS in the vicinity of the radar-BS
%The RIS-assisted ISAC system studied by \cite{jiang2021intelligent} considered a dual functional base station, a communication user, and a sensing target, where the signal-to-noise ratio (SNR) of radar is maximized while satisfying the communication user SNR constraints. 
After that, the authors in \cite{song2022joint} extended the above scenario to non-LoS (NLoS) multi-target sensing while assisting one single-antenna CU and maximizing the minimum beam directional gain pattern in the perceptual target direction while ensuring the constraints of the BS transmission power and the SNR of CU. A similar scenario was also considered in \cite{li2022dual}, where a dual-function BS simultaneously detects one single target and serves multiple CUs with the assistance of an RIS.
%further considered a single-user RIS-ISAC scenario with multiple targets to be detected, and proposed to maximize the minimum beampattern gain in several sensing directions under the transmit power constraint at the BS and the SNR constraint at the CU. 
Unlike the research of \cite{song2022joint} and \cite{li2022dual}, which focus on unique scenarios where there is no direct link between the BS and the target, \cite{zhang2022joint} and \cite{luo2023ris} studies more general application scenarios and considers multiple targets and multiple CUs.
The work of \cite{zhang2022joint}  studied the problem of  jointly optimizing mutual information (MI) for sensing and the sum data rate for communication. It alternately optimizes the beamforming matrices and the phase shifting matrix under the constraint of the transmission power of the BS. 
%In contrast, \cite{luo2023ris} focuses on improving target detection performance while ensuring CUs' QoS requirements and total transmission power budget.
In contrast, \cite{luo2023ris} focus on improving target detection performance while simultaneously ensuring the quality of service (QoS) requirements of the CUs and subjecting to the total transmission power budget.

%In \cite{liu2022joint}, extended the beamforming design to a multi-user RIS-ISAC system, and alternatively optimized the transmit beamformer, receive filter and reflection coefficients by maximizing the sum-rate of the UE under the radar SNR constraint. 

%For instance, the authors in \cite{wang2021joint} studied an IRS-assisted ISAC system with one base station (BS) and multiple communication users (CUs), in which only the communication was assisted by the RIS while the sensing was based on the direct LoS links. 

%Furthermore, the authors in \cite{jiang2021intelligent} considered a simplified ISAC setup with one BS, one single antenna CU, and one target, in which the signal-to-noise ratio (SNR) of radar is maximized while ensuring the SNR at the CU. 

%
%Note that all the above works only considered the communication of single-antenna CUs. Recently, DFRC beamforming designs were extended to multi-antenna CUs\cite{chen2022generalized}. However, to our knowledge, there is no relevant research on the ISAC system for RIS-assisted multi-antenna CUs.

Note that all the previous research works mentioned above only focused on the communication of single-antenna CUs. However, there have been recent advancements in DFRC beamforming designs, which now include considerations for multi-antenna CUs\cite{chen2022generalized}. Despite these developments, it is important to highlight that, to the best of our knowledge, there is currently no relevant research available on the topic of the ISAC system for RIS-assisted multi-antenna CUs.
%Besides, the system that we considered is dedicated to the detection of the target without LoS link and maximize the sum rate of multi-antenna CUs at the same time.  
%
%By deploying RIS appropriately, a solid virtual LoS link can be established in the system to achieve better communication/sensing performance.
In our study, we specifically focus on developing a system dedicated to detecting targets without LoS link, while simultaneously maximizing the sum rate of multi-user MIMO (MU-MIMO) communication. By strategically deploying the RIS, we can establish a virtual LoS link within the system, thereby significantly enhancing communication and sensing performance.
Against this background, the contributions of this article are summarized
%Motivated by the above observation, the main contributions of this paper are summarized as follows. 
as follows:
\begin{itemize}
	\item [1)] 
In this paper, we investigate an RIS-aided DFRC system, where the BS serves as a  radar to sense the target  with the assistance from the RIS and completes multi-user MIMO (MU-MIMO) communication.
%communicating with multiple multi-antenna communication users.
Our specific objective is to  maximize
the achievable sum rate of the CUs by jointly  optimizing beamforming matrix of the BS and the  phase shift matrix of the RIS while satisfying the constraint of  radar output SNR, the transmit power constraint of BS, and the unit modulus property of the reflecting coefficients of RIS. The main challenge in the problem lies in the coupling between optimized variables and the non-convex constraint of radar SNR.     
	\item [2)]
	To address the coupling problem of the optimization variables, we adopt a two-step approach. First, we reformulate the problem to an equivalent form that allows for easier decoupling of the variables. Next, we employ an alternating optimization (AO) algorithm to decouple the optimization variables and split this intractable problem into two subproblems. 
%	To solve the coupling problem of optimization variables, 
%	we first reformulate it into an equivalent one, then alternation optimization (AO)  is employed . 
	Given the  phase shift matrix, a penalty-based algorithm is proposed to deal with the non-convex radar SNR and transmit power constraints, including a two-layer iteration. The inner layer solves the penalized optimization problem, while the outer layer updates the penalty coefficient iteratively  to  ensure convergence. As for the given beamforming matrix of the BS, we apply majorization-minimization (MM) to transform the problem into a 
	quadratically constrained quadratic program (QCQP) problem and finally solve it by a semidefinite relaxation (SDR)-based algorithm.   
	\item [3)]
  Simulation results show that our proposed joint design for the RIS-assisted DFRC system can significantly help the DFRC system complete the sensing task.  At the same time, it can significantly enhance multi-antenna communication users' performance.  Specifically, compared to the benchmark scenario, our proposed joint design for the RIS-assisted DFRC system can improve the sum rate up to 5 $\mathrm{nat/s/Hz}$.
%   In an urban environment where surrounding buildings likely hinder BS-CUs path, our RIS-assisted DFRC system can always gain approximately 4 $\mathrm{bit/s/Hz}$.
\end{itemize}

The remainder of the paper is organized as follows. Section \ref{system} presents the system model and formulates the sum rate maximization problem. In Section \ref{algo}, we reformulate the original problem into a more tractable form, and the beamforming matrix of the BS and the passive reflecting coefficient of the RIS  are alternately optimized. In Section \ref{simulation}, simulation results are provided. Finally, our conclusions are offered in Section \ref{Conclusion}.

\textbf{Notations:}  Boldface lowercase and uppercase letters represent vectors and matrices, respectively. The transpose, conjugate and Hermitian operators are denoted by { $(\cdot)^{\rm{T}}$, $(\cdot)^{\rm{*}}$ and $(\cdot)^{\rm{H}}$}, respectively. The space of $M \times N$ complex matrices is represented by $\mathbb{C}^{M \times N}$, and the real part of $x$ is denoted by $\operatorname{Re}(x)$. $\operatorname{vec}(\mathbf{A})$ and $\operatorname{tr}(\mathbf{A})$  are vectorization  and the trace operators. The symbol $ \operatorname{unvec}{\left(\mathbf{q}\right)}$ means returning the elements of vector $ \mathbf{q} \in \mathbb{C}^{L^{2}\times 1}$ to a matrix of dimensions $ L \times L $. $\operatorname{diag}(\mathbf{A})$ is a vector composed of diagonal elements of matrix $\mathbf{A}$, while  $\operatorname{Diag}\left( \mathbf{a}\right) $  is a diagonal matrix having the entries of vector $ \mathbf{a}$ on its main diagonal.
%represent the diagonal stacking of a matrix into a vector, while  $\operatorname{Diag}\left( \mathbf{a}\right) $ defined as the operation of reconstructing elements in $\mathbf{a}$ into corresponding diagonal elements in a matrix.
$\mathbf{I}_{N}$ and $\mathbb{N}$ { represent the $N \times N$ identity matrix and set of natural numbers, respectively}. { $\mathbb{E}\left[  \cdot\right]   $ is the expectation operation.} { The $l_2$-norm}, Frobenius norm and absolute value operations are represented by $\|\cdot\|_{2}$, $\|\cdot\|_F$ and $|\cdot|$, respectively. The distribution of a circularly symmetric complex Gaussian random variable with zero mean and variance $\sigma^2$ is denoted by $\mathcal{C} \mathcal{N}\left(0, \sigma^2\right)$.  Finally, the Kronecker product and { Hadamard product} are represented by $\otimes$ { and $\odot$}, respectively.

\section{System Model and Problem Formulation}\label{system}

\subsection{System  Model}
\begin{figure}[]
	\centering
	\setlength{\belowcaptionskip}{6mm}
	\includegraphics[scale=0.28]{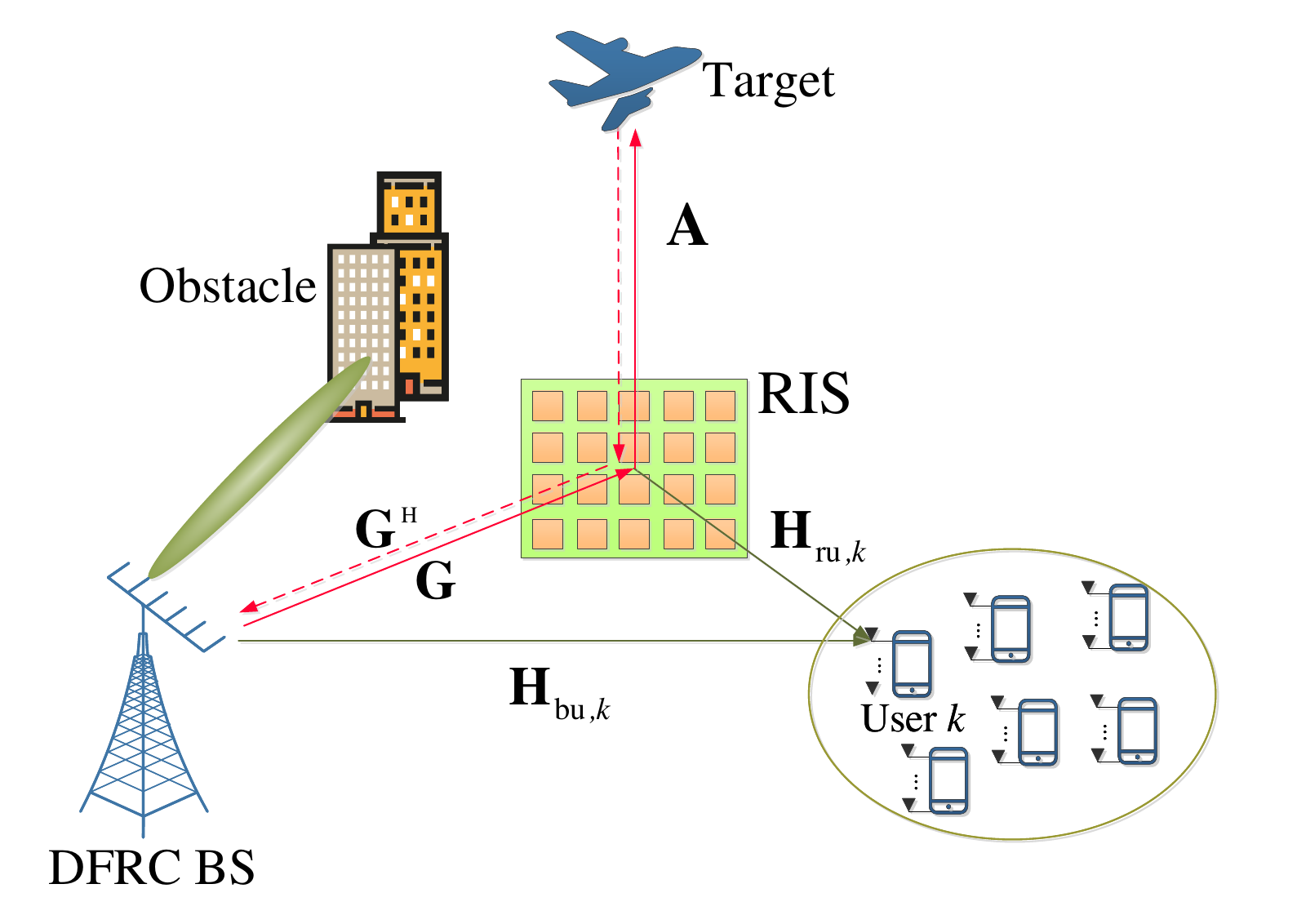}
	\captionsetup{font={small,stretch=1.25}}
	\caption{RIS-aided DFRC system with MU-MIMO communication.}\vspace{-0.8cm}
	\label{fig1}
\end{figure}

In Fig. \ref{fig1}, MU-MIMO communication is considered in an RIS-aided DFRC system. In the system, the BS serves a point target and $K$ multi-antenna users simultaneously. $N_{\rm{ t}}$ and  $N_{\rm{ r}}$  represent the transmit  and receiver antennas of the BS, respectively. We assume that there are $K$ CUs, each CU is equipped with $M_k$ antennas. We assume that there is no direct link between the radar and the target, necessitating the establishment of {virtual }LoS links with the support of an RIS panel comprising $L$ elements. Both the BS and the RIS are arranged as uniform linear arrays (ULAs).

%In this section, we consider an RIS-aided dual-function MU-MIMO communication and MIMO radar sensing system as shown in Fig. \ref{fig1}, consisting of an RIS, a radar target, a multi-antenna dual-function BS, and $K$ multi-antenna users. Without loss of generality, we assume that the BS is equipped with a uniform linear array (ULA) with $N_t$ transmit antennas and $N_r$ receiver antennas  to accomplish sensing tasks and communicate with multiple multi-antenna users. The $k$-th user has $M_k$ antennas with $k\in \{1,\cdots,K\}$, meaning multiple data streams $D_k$ can be transmitted between the $k$-th user and the BS. We assume there is no LoS between the radar and the target, so establishing virtual LoS links between them is necessary. As shown in Fig. \ref{fig1}, the DFRC system is aided by an RIS platform, modeled as an $L$-elements ULA. 

The dual-function BS transmits the signal ${\bf{x}}\in \mathbb {C}^{N_{\rm{ t}} \times 1}$ as
\begin{equation}\label{transmit_signal}
	{{\bf{x}}} = \sum\limits_{k = 1}^K {{\bf{B}}_{k}{\bf{d}}_{k}},
\end{equation}
where ${\bf{B}}_k \in \mathbb{C}^{N_{\rm{ t}} \times D_k}$ {is the beamforming matrix of CU $k$}. The independent random variable ${\bf{d}}_k$ denotes the  data vector for user $k$, satisfying $ \mathbf{d}_{k} \sim \mathcal{C N}\left(\mathbf{0}, \mathbf{I}_{D_{k}}\right) $.  $D_{k}$ is the dimension of ${\bf{d}}_k$, which should not  be larger than  $\min \left\{N_{\rm{ {t}}}, M_{k}\right\}$. Therefore, the transmit power satisfies the following constraint:
%where ${\bf{B}}_k \in \mathbb{C}^{N_t \times D_k}$ and ${\bf{d}}_k \in \mathbb{C}^{D_k \times 1}$ are the transmit beamforming matrices and the data vectors of the $k$-th user with $ D_{k} \leq \min \left\{N_{t}, M_{k}\right\} $, respectively. ${\bf{d}}_k$ is assumed to satisfy Gaussian distributed with $ \mathbf{d}_{k} \sim \mathcal{C N}\left(\mathbf{0}, \mathbf{I}_{D_{k}}\right) $ and the data vectors of different users are assumed to be independent, satisfying $\mathbb{E}\left[ {{{\bf{d}}_{i}}{{{\bf{d}}_{j}^H} }} \right] = {\bf{0}}$, ${\rm{for}}\  \{i\} \ne \{j\}$. 
\begin{equation}\label{transmit_power}
\mathbb{E}\left[ \Vert \mathbf{x} \Vert_{2}^{2}\right] =\sum_{k=1}^{K} \operatorname{tr}\left(\mathbf{B}_{k} \mathbf{B}_{k}^{\rm{H}}\right) \leq P_{0},
\end{equation}
where $ P_{0} $ is the maximum transmit power.

%Let ${\bf{G}} \in \mathbb{C}^{L \times N_t}$ denote the channel between the dual-function BS and the RIS, ${\bf{H}}_{bu,k}\in \mathbb{C}^{M_k \times N_t}$ denote that from the dual-function BS to user $k$, and ${\bf{H}}_{ru,k}\in \mathbb{C}^{M_k \times L}$ denote that from the RIS to user $k$, $k \in \{1,\cdots,K\}$. The signal can reach the communication receivers either through a direct path or by reflection from the RIS.

Consider $\mathbf{G} \in \mathbb{C}^{L \times N_{\rm{ t}}}$ as the channel matrix representing the connection between the dual-function BS and the RIS. Additionally, define $\mathbf{H}_{\mathrm{bu},k} \in \mathbb{C}^{M_k \times N_{\rm{ t}}}$ as the channel  characterizing the link from the  BS to CU $k$, and $\mathbf{H}_{\mathrm{ru},k} \in \mathbb{C}^{M_k \times L}$ as the channel  representing the connection from the RIS to CU $k$. We assume that the above channels' channel state information (CSI) is perfectly known at the DFRC BS by applying effective channel estimation methods\cite{zhou2022channel}.  The signal received by the communication receivers can be acquired either directly or via reflection from the RIS. Accordingly, the signal reception at the $k$-th CU can be  described as

\begin{equation}\label{user_k_signal}
{\bf{y}}_k=\left( {\bf{H}}_{\mathrm{bu},k} + {\bf{H}}_{\mathrm{ru},k} {\mathbf{\Theta}}^{\rm{H}} {\bf{G}} \right) 	{{\bf{x}}} + {\bf{z}}_{k} ,
\end{equation}
where $ {\mathbf{\Theta}}  = {\operatorname{Diag}}\left(  {\boldsymbol{\vartheta}}\right) $ is the phase shift matrix of the RIS and {${\boldsymbol{\vartheta}} =  \left[ {{\vartheta} _1} , \cdots ,{{\vartheta} _l} , \cdots ,{{\vartheta} _L} \right]^{\mathrm{T}} $} with $\left| {{\vartheta} _l} \right| = 1$, $l\in \mathcal{L}, \mathcal{L}\triangleq\left\lbrace 1,2,\dots,L\right\rbrace $. The vector ${\bf{z}}_{k} \sim \mathcal{C N}{\left(\mathbf{0}, \sigma^{2}\mathbf{I}_{M_k}\right)} $ denotes the additive white Gaussian noise (AWGN) with variance  $\sigma^{2}$. We can define  ${\bf{H}}_{k} \triangleq {\bf{H}}_{\mathrm{bu},k} + {\bf{H}}_{\mathrm{ru},k} {\mathbf{\Theta}}^{\rm{H}} {\bf{G}}$, as the representation of the equivalent  channel  from the BS to CU $k$. Thus, by substituting (\ref{transmit_signal}) into (\ref{user_k_signal}), ${\bf{y}}_k $ is { rewritten} as
\begin{equation}\label{user_k_signal_H_K}
{\bf{y}}_k={\bf{H}}_{k} \sum\limits_{i = 1}^K {{\bf{B}}_{i}{\bf{d}}_{i}} + {\bf{z}}_{k}.
\end{equation}

{Let} ${\bf{J}}_{k} = \sigma^{2}{\bf{I}}_{M_k} + \sum\limits_{i = 1,i \neq k}^K  {\bf{H}}_{k}    {\bf{B}}_{i} {\bf{B}}_{i}^{\rm{H}}  {\bf{H}}_{k}^{\rm{H}}$ { contain the noise and the interference,}  the achievable  sum rate is expressed as
\begin{equation}\label{sum_data_rate}
	{R} =  \sum\limits_{k = 1}^K{\log}\left| {\bf{I}}_{M_k} + {\bf{H}}_{k} {\bf{B}}_{k}{\bf{B}}_{k}^{\rm{H}}  {\bf{H}}_{k}^{\rm{H}}   {\bf{J}}_{k}^{-1} \right|.
\end{equation}
%
%Based on the assumption of Gaussian distributed data vectors for each user, the achievable data rate  of the $k$-th user is given by
%\begin{equation}\label{data_rate_k}
%	{R_{k}} = {\log}\left| {\bf{I}} + {\bf{B}}_{k}^{\rm{H}}  {\bf{H}}_{k}^{\rm{H}} {\bf{J}}_{k}^{-1}  {\bf{H}}_{k} {\bf{B}}_{k} \right|, \quad {\forall} k
%\end{equation}
%where
%\begin{equation}\label{interference_matrix} 
%	{\bf{J}}_{k} = \sigma^{2}{\bf{I}} + \sum\limits_{i = 1,i \neq k}^K  {\bf{H}}_{k}    {\bf{B}}_{i} {\bf{B}}_{i}^{\rm{H}}  {\bf{H}}_{k}^{\rm{H}}.
%\end{equation}
%Then, the sum rate of the MU-MIMO communication can be given by
%\begin{equation}\label{sum_data_rate}
%	{R} =  \sum\limits_{k = 1}^K{\log}\left| {\bf{I}} + {\bf{B}}_{k}^{\rm{H}}  {\bf{H}}_{k}^{\rm{H}} {\bf{J}}_{k}^{-1}  {\bf{H}}_{k} {\bf{B}}_{k} \right|.
%\end{equation}

%In an urban environment, the presence of surrounding buildings often obstructs the LoS path, which in turn has an impact on the radar sensing capabilities of the system.
%When there is no direct path between the dual-function BS and the target, the dual-function BS transmits a signal which is then reflected by the RIS. The reflected signal travels towards the target and upon reaching it, the target's echo is received by the BS receiver via the RIS. As a result, the received signal can be expressed as follows:

{ In urban environments, the presence of surrounding structures often leads to the obstruction of the LoS path, thereby affecting the radar sensing capabilities of the system.} When a direct path between the dual-function BS and the target is absent, the dual-function BS transmits a signal, which is subsequently reflected by the RIS. This reflected signal propagates toward the target, and upon its arrival, the target's echo is received by the BS receiver through the RIS. Consequently, {the received echo signal at the BS} can be mathematically represented as
\begin{eqnarray}\label{radar_signal}
	{\bf{y}}_0=
%	{\bf{G}}^{\rm {H}} {\mathbf{\Theta}}{{\mathbf{a}}}\left( \theta_0 \right) {{{\mathbf{a}}}^{H}}\left( \theta_0 \right){\mathbf{\Theta}}^{\rm{H}} {\bf{G}} 	{{\bf{x}}} + {\bf{z}}_{0} \\
	{\bf{G}}^{\rm {H}} {\mathbf{\Theta}} {\bf{A}} {\mathbf{\Theta}}^{\rm{H}} {\bf{G}} 	{{\bf{x}}} + {\bf{z}}_{0} ,
\end{eqnarray}
where $ {\bf{z}}_{0} $ models the AWGN satisfying $ \mathcal{C N}\left({\bf{0}}, \sigma_\mathrm{r}^{2}\mathbf{I}_{N_{\mathrm{r}}}\right) $ at the radar receiver with variance  $\sigma_{\mathrm{r}}^{2}$. The matrix ${\bf{A}}\in {\mathbb{C}}^{L \times L}$  is the target response matrix of the RIS. 
Assuming that RIS can be regarded as a monostatic MIMO radar \cite{jiang2021intelligent}, the point target response matrix can be defined as 
\begin{equation}\label{A}
	{\bf{A}} ={{\bf{a}}}\left( \theta_0 \right) {{{\bf{a}}}^{\mathrm{H}}}\left( \theta_0 \right) ,
\end{equation}
where $\theta_0$ is the direction of the target with respect to (w.r.t.) the RIS.%${\bf{a}}(\theta_0)=\left[1, \cdots, e^{-j 2 \pi d \left(L-1\right)\sin \theta_0 / \lambda}\right]^{\mathrm{T}} \in \mathbb{C}^{L \times 1}$ is the steering vector of the RIS, where $d$ and $\lambda$ denote the interelement spacing and the signal wavelength, respectively.
%where $ {\bf{z}}_{0} \in \mathcal{C}^{N_{r} \times 1} $ models the additive white Gaussian noise (AWGN) satisfying $ \mathcal{C N}\left({\bf{0}}, \sigma_\mathrm{r}^{2}\mathbf{I}_{N_{r}}\right) $ at the radar receiver and $ \sigma_r^{2} $ is the average noise power per radar receiver antenna. The matrix ${\bf{A}}\in {\mathbb{C}}^{L \times L}$  is the target response matrix of the RIS. Considering a point target located at angle $\theta_0$ w.r.t the  RIS, the expression of the point  target response matrix $\bf{A}$ can be defined as follow under the assumption of  considering RIS as a monostatic MIMO radar:
The vector ${\bf{a}}\left(\theta_0\right)$ is the steering vector associated with the RIS, defining by
\begin{equation}\label{a}
  {\bf{a}}(\theta_0)=\left[1, \cdots, e^{-j 2 \pi d \left(L-1\right)\sin \theta_0 / \lambda}\right]^{\mathrm{T}} ,
 \end{equation}
where $d$ and $\lambda$ denote the {element }spacing and the signal wavelength, respectively.

Following this, we can describe the output received by the radar receiver as 
\begin{equation}\label{radar_SCNR}
	r=\mathbf{w}^{\rm{H}} \mathbf{y}_{0}=\eta\mathbf{w}^{\rm{H}}\mathbf{G}^{\mathrm{H}} \mathbf{\Theta A} \boldsymbol{\Theta}^{\mathrm{H}} \mathbf{G x}+\mathbf{w}^{\rm{H}}\mathbf{z}_{0},
\end{equation}
where $\eta$ is the channel gain and $\mathbf{w} \in \mathbb{C}^{N_{\rm{ t}} \times 1}$ denotes the receive beamforming vector. In order to maximize the  radar SNR, minimum variance distortionless response (MVDR) algorithm can be utlized \cite{capon1969high}, which gives the optimal $\mathbf{w}^{\star}$ as
%\begin{eqnarray}
\begin{equation}\label{omega}
	\mathbf{w}^{\star} = {\mathop{\operatorname{argmax}}\limits_{ \mathbf{w} }    }\frac{\left|\mathbf{w}^{\rm{H}} \mathbf{V}\mathbf{x}\right|^{2}}{\mathbf{w}^{\rm{H}}\mathbf{w}}  =\beta \mathbf{V}\mathbf{x},
\end{equation}
where $\beta$ represents a constant and ${\bf{V}} \triangleq  \eta \mathbf{G}^{\mathrm{H}} \mathbf{\Theta A} \boldsymbol{\Theta}^{\mathrm{H}} \mathbf{G}$. Thus, the corresponding  radar SNR can be computed as
\begin{eqnarray}
 	 \gamma & =&{ \mathbb{E}}\left[\frac{\left|\mathbf{w}^{\rm{H}}\bf{V}\mathbf{x}\right|^{2}}{ \sigma_{\mathrm{r}}^{2}\mathbf{w}^{\rm{H}}\mathbf{w}}\right] \nonumber
 	 \\ 
 	 &{\stackrel{\left( a\right) }{=}}&\mathbb{E}\left[\mathbf{x}^{\rm{H}}{\bf{V}}^{\rm{H}}{\bf{V}} \mathbf{x}\right]/ \sigma_\mathrm{r}^{2}\nonumber 
 	 \\ 
 	 &\stackrel{(b)}{=}&\sum_{k=1}^{K} \operatorname{tr}\left({\bf{V}}^{\rm{H}}{\bf{V}}\mathbf{B}_{k} \mathbf{B}_{k}^{\rm{H}}\right)/ \sigma_\mathrm{r}^{2},
\end{eqnarray}
where Step (a) corresponds to the determination of the optimal receive beamforming vector ${\bf{w}}^{\star}$, as described in equation (\ref{omega}). Meanwhile, Step (b) results from the mathematical expectation ${ { \mathbb{E}}\left[{\bf{xx}}^{\rm{H}}\right] }= \sum_{k=1}^{K} \mathbf{B}_{k}{\bf{B}}_{k}^{\rm{H}}$.

\subsection{Problem Formulation}

In this paper, we aim to jointly optimize the transmit beamforming matrix ${\bf{B}}_{k}$ and the phase shift  matrix $\bf{\Theta}$ to maximize the sum rate of the MU-MIMO communication. Specifically, the maximization problem  of the MU-MIMO communication sum rate is formulated as

%In this paper, we aim to maximize the sum rate of the MU-MIMO communication of all the users by jointly optimizing the transmit beamforming matrices ${\bf{B}}_{k}$ at the BS and the phase shifts  matrix $\bf{\Theta}$ at the RIS. Specifically, the sum rate of the MU-MIMO communication maximization problem is formulated as follows:
\begin{subequations}\label{comm_cen}
	\begin{align}
		\mathop {\max }\limits_{{ \left\lbrace \mathbf{B}_k\right\rbrace} ,\mathbf{\Theta}}\quad &R
		\\
		\qquad\ \textrm{s.t.}\quad
		&\mathbf{\gamma} \geq  \gamma_0, \label{SNR}
		\\
		& \sum_{k=1}^K \operatorname{tr}\left(\mathbf{B}_k \mathbf{B}_k^{\rm{H}}\right) \leq P_0,  \label{Power}
		\\
		&\left| {{\vartheta} _l} \right| = 1, l\in \mathcal{L},\label{phase}
	\end{align}
\end{subequations}
where $\gamma_0$ is the required minimum SNR to guarantee the radar sensing performance. The complexity of addressing this optimization problem can be attributed to the intricate coupling that exists between the transmit beamforming matrix ${\bf{B}}_{k}$ and the phase shift matrix $\bf{\Theta}$. Furthermore, the constraints delineated in  (\ref{phase}) exacerbate the challenge.

 \section{The transmit beamforming and  RIS phase joint design method}\label{algo}
%Within this section, we start by transforming the original problem into an equivalent one by relating data rate and weighted minimum mean-square error (WMMSE). This approach allows us to lay the foundation for subsequent optimization steps.
In this section, we commence by reformulating Problem (\ref{comm_cen}) into an equivalent one through the association between data rate and the principle of weighted minimum mean-square error (WMMSE). This approach allows us to lay the foundation for subsequent optimization steps. Subsequently, we utilize the AO method to address the reformulated problem, acknowledging the coupling among optimization variables. The method facilitates the decomposition of the intricate problem into two noticeable subproblems, each dedicated to optimizing specific variables, thereby enhancing the tractability of the overall optimization procedure. To address the optimization problem associated with transmit beamforming matrix, we implement a penalty-based algorithm. Concurrently, we employ the MM algorithm to reformulate the quartic non-convex constraint and subsequently utilize the SDR method for phase shift matrix optimization.
%Simultaneously, we utilize the MM  algorithm to transform the  non-convex constraint into a more manageable form. Subsequently, we apply the SDR method to optimize the phase shift matrix. 
Via this approach, we can address the coupling between  ${\bf{B}}_{k}$ and the phase shift matrix $\bf{\Theta}$ and handle the non-convex constraints effectively, ultimately optimizing the system performance.

%A penalty-based algorithm is utilized for solving the optimization of transmit beamforming matrices problem. At the same time, we use MM algorithm to trans the quartic non-convex  constraint into a more tractable form, and then the SDR method is used to solve the optimization problem of the phase shifts  matrix. 
\subsection{Reformulation of the Original Problem}
Without loss of generality, we consider an arbitrary CU $k$ and show how to reformulate the corresponding problem. We assume a linear decoding matrix $\mathbf{U}_k\in \mathbb{C}^{M_k\times D_k}$, which allows us to compute the estimated signal vector ${{{\bf{\hat d}}}_{k}} $ for CU $k$ as 
\begin{equation}\label{decode}
	{{{\bf{\hat d}}}_{k}} = {\bf{U}}_{k}^{\rm{H}}{{\bf{y}}_{k}},\forall k.
\end{equation}

{ Then}, we can calculate the mean-square error (MSE) matrix for the $k$-th CU as
\begin{eqnarray}\label{MSE_k}
	%\begin{aligned}
	\mathbf{E}_{k} &=& \mathbb{E}\left[\left( {\bf{\hat d}}_{k} - {\bf{d}}_k \right) {\left( {\bf{\hat d}}_{k} - {\bf{d}}_k \right) 	}^{\rm{H}}	\right] \nonumber
	\\
	&=& {\bf{U}}_{k}^{\rm{H}}{\bf{H}}_{k}\left( \sum\limits_{i = 1}^K  {{\bf{B}}_{i}}{\bf{B}}_{i}^{\rm{H}}\right) {\bf{H}}_{k}^{\rm{H}}{\bf{U}}_{k}
	-{\bf{U}}_{k}^{\rm{H}}{\bf{H}}_{k}{{\bf{B}}_{k}}\nonumber
	\\
	&\quad&-{\bf{B}}_{k}^{\rm{H}}{\bf{H}}_{k}^{\rm{H}}{\bf{U}}_{k}
	+\sigma^2{\bf{U}}_{k}^{\rm{H}}{\bf{U}}_{k}+{\bf{I}}_{D_k}\label{MSE_k_2}.
	%\end{aligned}
\end{eqnarray}

Subsequently, the expression for $R$ in (\ref{sum_data_rate}) can be transformed as  
%by introducing a set of auxiliary matrices $ $
%the sum of the weighted MSE can be defined as
\begin{equation}\label{sum_MSE}
{E}=\sum_{k=1}^{K} \left( \log\left| \mathbf{W}_{k}\right| -\operatorname{tr}\left(\mathbf{W}_{k} \mathbf{E}_{k}\right) + D_k\right) ,
\end{equation}
where ${\bf{W}}_{k}\in \mathbb{C}^{{D_{k}\times D_k}}$ represents the auxiliary  matrix for the $k$-th CU and $D_k$ is the constant term. 

Hence, by ignoring the constant term of (\ref{sum_MSE}), we can  reformulate Problem (\ref{comm_cen}) as follows\cite{shi2011iteratively}:
\begin{subequations}\label{Reformulation}
	\begin{align}
		\mathop {\min }\limits_{\left\lbrace \mathbf{B}_k , {\bf{W}}_k ,{\bf{U}}_k\right\rbrace,\mathbf{\Theta} }\quad &\sum_{k=1}^{K} \left(\operatorname{tr}\left(\mathbf{W}_{k} \mathbf{E}_{k}\right)  -  \log\left| \mathbf{W}_{k}\right| \right) \label{16a}
		\\
	 \textrm{s.t.}\qquad\quad
		&(\ref{SNR}),(\ref{Power}),(\ref{phase}),
%		(\ref{Power}),(\ref{SNR}),(\ref{phase})	.	
	\end{align}
\end{subequations}
due to the challenges posed by the coupled variables in Problem (\ref{Reformulation}), attaining the optimal solution is an awkward task. In order to address this issue, the most widely used and practical approach is to employ an AO mechanism. Specifically, we optimize a specific group of variables while maintaining the others fixed to minimize the objective function (\ref{16a}).
%Specifically, we minimize the objective function in (\ref{Reformulation}) by alternately optimizing one set of optimization variables while keeping the other variables fixed. 
In Problem (\ref{Reformulation}), it is discernible that matrices $ \mathbf{U}_{k} $ and ${\bf{W}}_k$ are exclusively associated with the objective function. Consequently, we commence by setting the first-order derivative of ${E}$ w.r.t. $\mathbf{U}_{k}$ to zero while maintaining the other three variables fixed. This approach enables us to deduce the optimal solution for $\mathbf{U}_{k}$ as \cite{pan2020intelligent,peng2021multiuser}
\begin{equation}\label{decode_U}
	\mathbf{U}_{k}^{\mathrm{opt}}=\left({\bf{H}}_{k} \sum_{i=1}^K \mathbf{B}_i \mathbf{B}_i^{\rm{H}} {\bf{H}}_{k}^{\rm{H}} + \sigma^{2}\mathbf{I}_{M_k}\right)^{-1}{\bf{H}}_{k}{\mathbf{B}_k}.
\end{equation}
%In the following, we can derive the optimal solution for $ \mathbf{U}_{k} $ and ${\bf{W}}_k$ when the other matrices are fixed. For given values of $ \boldsymbol{\Theta}$, $ {\mathbf{W}} $, and $ {\mathbf{B}} $, we can set the first-order derivative of $ {E} $ w.r.t. $ \mathbf{U}_{k} $ to zero, which gives the optimal $ \mathbf{U}_{k} $:
%\begin{equation}\label{decode_U}
%\mathbf{U}_{k}=\left({\bf{H}}_{k} \sum_{i=1}^K \mathbf{B}_i \mathbf{B}_i^{\rm{H}} {\bf{H}}_{k}^{\rm{H}} + \sigma^{2}\mathbf{I}\right)^{-1}{\bf{H}}_{k}{\mathbf{B}_k}.
%\end{equation}

Similarly, upon fixing the values for the other three matrices except $\mathbf{W}_{k} $, we can deduce the optimal solution for matrix $\mathbf{W}_{k}$ as
\begin{equation}\label{Weight_W}
\mathbf{W}_{k}^{\mathrm{opt}}=\mathbf{E}_{k}^{-1}.
\end{equation}

\subsection{Optimize the Transmit Beamforming Matrix $\mathbf{B}_k$} \label{opt_b}
Our primary focus is on optimizing the transmit beamforming matrix $\left\lbrace \mathbf{B}_k\right\rbrace $, while keeping $\boldsymbol{\Theta}$, $\left\lbrace {\mathbf{W}}_k\right\rbrace $, and $\left\lbrace {\mathbf{U}}_k\right\rbrace $ fixed. First of all, 
we substitute $\mathbf{E}_k$ into (\ref{sum_MSE}) and remove the constant terms, then the transmit  beamforming matrix optimization problem is formulated as
%Then, we introduce  auxiliary variable $\mathbf{X}_k, \mathbf{Y}_k $ and define $ \mathbf{X}_k =\mathbf{B}_{k} ,\mathbf{Y}_k = \mathbf{V}\mathbf{B}_{k}$. %Subsequently, we use these {as penalty terms which are added to the} objective function, yielding the following penalty-based optimization problem:
%Following this, we incorporate these terms as penalties added to the objective function, resulting in the following penalty-based optimization problem:
%In this subsection, we focus our attention on optimizing the transmit beamforming matrices $\mathbf{B}_k$, while fixing $\boldsymbol{\Theta}$, $ {\mathbf{W}}$, and $ {\mathbf{U}} $. By substituting $\mathbf{E}_k$ into (\ref{sum_MSE}) and removing the constant terms, the  transmit  beamforming matrices optimization problem is formulated as
\begin{subequations}\label{otp_bk}
	\begin{flalign}
		\mathop {\min }\limits_{\left\lbrace \mathbf{B}_k\right\rbrace }\quad &-\sum_{k=1}^K\operatorname{tr}\left(\mathbf{W}_k\mathbf{U}_k^{\rm{H}}\mathbf{H}_k\mathbf{B}_k\right)\nonumber
		\\
		&+\sum_{k=1}^K\operatorname{tr}\left(\mathbf{B}_k^{\rm{H}}\sum_{m=1}^{K}\mathbf{H}_m^{\rm{H}}\mathbf{U}_m\mathbf{W}_m\mathbf{U}_m^{\rm{H}}\mathbf{H}_m\mathbf{B}_k\right)\nonumber
		\\
		&-\sum_{k=1}^K\operatorname{tr}\left(\mathbf{W}_k\mathbf{B}_k^{\rm{H}}\mathbf{H}_k^{\rm{H}}\mathbf{U}_k\right)
		\label{B_obj}
		\\
		\qquad\  \textrm{s.t.}\quad
		&	(\ref{SNR}),(\ref{Power}).
%		\sum_{k=1}^K \operatorname{tr}\left(\mathbf{B}_k \mathbf{B}_k^{\rm{H}}\right) \leq P_0\label{opt_bk_B} ,
%		\\
%		&\mathbf{\gamma} =\sum_{k=1}^{K} \operatorname{tr}\left({\bf{V}}^{\rm{H}}{\bf{V}}\mathbf{B}_{k} \mathbf{B}_{k}^{\rm{H}}\right)\geq \sigma^{2}\gamma_0.
	\end{flalign}
\end{subequations}

Problem (\ref{otp_bk}) poses a substantial challenge for resolution, primarily attributed to the non-convex constraints (\ref{SNR}). To address the non-convex constraints, a penalty-based algorithm is utilized\cite{hua2022joint}. Specifically, we first introduce  auxiliary variables $\mathbf{X}_k, \mathbf{Y}_k $ and define $ \mathbf{X}_k =\mathbf{B}_{k} ,\mathbf{Y}_k = \mathbf{V}\mathbf{B}_{k}$. Then  Problem (\ref{otp_bk}) can be rewritten as 
\begin{subequations}\label{auxiliary_X_Y}
	\begin{align}
		\mathop {\min }\limits_{\left\lbrace \mathbf{B}_k, \mathbf{X}_k , \mathbf{Y}_k\right\rbrace  } 
		&-\sum_{k=1}^K\operatorname{tr}\left(\mathbf{W}_k\mathbf{U}_k^{\rm{H}}\mathbf{H}_k\mathbf{B}_k\right)\nonumber
		\\
		&+\sum_{k=1}^K\operatorname{tr}\left(\mathbf{B}_k^{\rm{H}}\sum_{m=1}^{K}\mathbf{H}_m^{\rm{H}}\mathbf{U}_m\mathbf{W}_m\mathbf{U}_m^{\rm{H}}\mathbf{H}_m\mathbf{B}_k\right)\nonumber
		\\
		&-\sum_{k=1}^K\operatorname{tr}\left(\mathbf{W}_k\mathbf{B}_k^{\rm{H}}\mathbf{H}_k^{\rm{H}}\mathbf{U}_k\right)
		\label{B_obj20}
%		\qquad (\ref{B_obj})
		\\
		\textrm{s.t.}\quad
		&\quad\sum_{k=1}^K \operatorname{tr}\left(\mathbf{X}_k\mathbf{X}_k^{\rm{H}}\right) \leq P_0,\label{X_Con}
		\\
		&\quad\sum_{k=1}^{K} \operatorname{tr}\left(\mathbf{Y}_k\mathbf{Y}_k^{\rm{H}}\right)\geq \sigma_\mathrm{r}^{2} \gamma_0,\label{Y_Con}
		\\
		&\quad\mathbf{X}_k = \mathbf{B}_{k} ,\mathbf{Y}_k = \mathbf{V}\mathbf{B}_{k},\quad k = 1, \cdots ,K \label{auxiliary}.
	\end{align}
\end{subequations}

Subsequently, we incorporate (\ref{auxiliary}) as penalty terms which are introduced into (\ref{B_obj20}), resulting in the following penalty-based optimization problem
\begin{subequations}\label{Problem_with_auxi_in_obj}
	\begin{align}
		\mathop {\min }\limits_{ {\left\lbrace \mathbf{B}_k  , \mathbf{X}_k ,\mathbf{Y}_k\right\rbrace } } &-\sum_{k=1}^K\operatorname{tr}\left(\mathbf{W}_k\mathbf{U}_k^{\rm{H}}\mathbf{H}_k\mathbf{B}_k\right)\nonumber	-\sum_{k=1}^K\operatorname{tr}\left(\mathbf{W}_k\mathbf{B}_k^{\rm{H}}\mathbf{H}_k^{\rm{H}}\mathbf{U}_k\right)\nonumber
		\\
		&+\sum_{k=1}^K\operatorname{tr}\left(\mathbf{B}_k^{\rm{H}}\sum_{m=1}^{K}\mathbf{H}_m^{\rm{H}}\mathbf{U}_m\mathbf{W}_m\mathbf{U}_m^{\rm{H}}\mathbf{H}_m\mathbf{B}_k\right)\nonumber
		\\	
		&+\frac{1}{2\rho^{\left[ t\right] }}\left( \sum_{k=1}^K \Vert {\mathbf{X}_k} - \mathbf{B}_{k}\Vert_{F}^{2}+ \sum_{k=1}^K \Vert {\mathbf{Y}_k} -  \mathbf{V}\mathbf{B}_{k}\Vert_{F}^{2}\right) \label{obj_value}
		\\
		 \textrm{s.t.}\quad\
		 &(\ref{X_Con}),(\ref{Y_Con}),
%		 &\sum_{k=1}^K \operatorname{tr}\left(\mathbf{X}_k\mathbf{X}_k^{\rm{H}}\right) \leq P_0,
%		 \\
%		 &\mathbf{\gamma} =\sum_{k=1}^{K} \operatorname{tr}\left(\mathbf{Y}_k\mathbf{Y}_k^{\rm{H}}\right)\geq \sigma_r^{2} \gamma_0, 
%%		&\sum_{k=1}^K \operatorname{tr}\left(\mathbf{X}_k\mathbf{X}_k^{\rm{H}}\right) \leq P_0,
%		\\
%		&\mathbf{\gamma} =\sum_{k=1}^{K} \operatorname{tr}\left(\mathbf{Y}_k\mathbf{Y}_k^{\rm{H}}\right)\geq \sigma^{2}\gamma_0,
	\end{align}
\end{subequations}
where $\rho^{\left[ t\right] }\left(\rho^{\left[ t\right] } > 0 \right) $  is the penalty coefficient associated with constraint (\ref{auxiliary}) in the $t$-th iteration. During the whole iteration,  We aim to encourage the optimization process to satisfy  (\ref{auxiliary}) closely as possible while maintaining the goal of minimizing the objective function. To attain this objective, in the outer iteration, we  decrease the penalty coefficient's value, driving the coefficient $\frac{1}{2 \rho^{\left[ t\right] }}$  toward infinity. Ultimately, through the optimization process, the equality constraint is  achieved. Essentially, the method offers a way to strike a balance in  optimization, aligning it with the dual goals of achieving the  original objective function and complying with the equality constraints.
%where $\rho\left(\rho > 0 \right) $  represents the penalty coefficient used to
%penalize the violation of the equality in constraint (\ref{auxiliary}). By
%gradually decreasing the value of $ \rho$ over outer layer iterations 
%as $\rho \rightarrow 0$,  it follows that $1 /(2 \rho) \rightarrow \infty$. As such, the equality in (\ref{auxiliary}) is guaranteed by the optimal solution to Problem (\ref{Problem_with_auxi_in_obj}). 

In the inner layer of the penalty-based algorithm, the penalty coefficient $\rho^{\left[ t\right] }$ remains constant. However, the problem concerning these three optimization variables is still non-convex. Fortunately, it can be decomposed into two subproblems: one for optimizing the beamforming matrix and the other for optimizing the auxiliary matrix, respectively. These two subproblems are alternately solved until  convergence.

In the following sections, we provide a comprehensive exposition of the procedures involved in both the outer and inner optimization layers within the penalty-based algorithm.

%With fixed $\rho$, it can be seen that Problem (\ref{Problem_with_auxi_in_obj}) is still non-convex. To tackle this non-convex optimization problem, we divide all the optimization variables into two blocks in the inner layer, namely, 1) transmit beamforming $\left\{\mathbf{B}_{k}\right\}, $2) auxiliary variables $\left\{ \mathbf{X}_{k}, \mathbf{Y}_k\right\}$, and then alternately optimize each block, until the convergence is achieved.
\subsubsection{ Outer Layer Update}
In the outer layer, the penalty coefficient $\rho^{\left[ t\right] }$ in the $t$-th iteration can be updated as
\begin{equation}\label{rho}
	\rho^{\left[ t\right] }=c \rho^{\left[ t-1\right] },
\end{equation}
where $c \left( 0<c<1\right) $ represents the updated step size. Subsequently, without loss of generality, we  take the $t$-th outer layer update as an example to specifically introduce the inner layer optimization.

\subsubsection{Inner Layer Optimization}
%In this subsection, we elaborate on how to solve the above two subproblems. In particular, we obtain a closed-form and a semi-closed-form solution to these  two subproblems.
%For given transmit beamforming matrix $\left\{ \mathbf{B}_{k}\right\} $ and the auxiliary matrix $ \left\{\mathbf{X}_k\right\}$, $ \left\{\mathbf{Y}_k\right\}$ can be optimized by solving the following subproblem ({only items related to $\mathbf{Y}_k$} are kept)

\paragraph{With fixed $\left\{\mathbf{X}_k, \mathbf{Y}_k \right\}$, optimize $\left\{ \mathbf{B}_{k}\right\}$}For given auxiliary variables $\left\{ \mathbf{X}_{k}, \mathbf{Y}_k\right\} $, the subproblem associated with the optimization of transmit beamforming can be formulated as
\begin{equation}\label{Fix_X_Y}		
	\begin{aligned}
		\begin{split}
		\mathop {\min }\limits_{ \left\lbrace {\mathbf{B}_k }\right\rbrace   } &-\sum_{k=1}^K\operatorname{tr}\left(\mathbf{W}_k\mathbf{U}_k^{\rm{H}}\mathbf{H}_k\mathbf{B}_k\right)-\sum_{k=1}^K\operatorname{tr}\left(\mathbf{W}_k\mathbf{B}_k^{\rm{H}}\mathbf{H}_k^{\rm{H}}\mathbf{U}_k\right)
		\\
		&+\sum_{k=1}^K\operatorname{tr}\left(\mathbf{B}_k^{\rm{H}}\sum_{m=1}^{K}\mathbf{H}_m^{\rm{H}}\mathbf{U}_m\mathbf{W}_m\mathbf{U}_m^{\rm{H}}\mathbf{H}_m\mathbf{B}_k\right)
		\\
		&+\frac{1}{2\rho^{(t)}}\left( \sum_{k=1}^K\Vert {\mathbf{X}_k} - \mathbf{B}_{k}\Vert_{F}^{2}+ \sum_{k=1}^K\Vert {\mathbf{Y}_k} -  \mathbf{V}\mathbf{B}_{k}\Vert_{F}^{2}\right) .
		\end{split}
	\end{aligned}
\end{equation}

Given that Problem (\ref{Fix_X_Y}) is an unconstrained convex quadratic minimization problem, we can find the optimal solution by utilizing the first-order optimality conditions. Specifically, setting the first-order derivative of  (\ref{Fix_X_Y}) w.r.t. $\mathbf{B}_{k}$ to zero, we obtain the optimal solution for  $\mathbf{B}_{k}$ as 
%Specifically, by taking the first-order
%derivative of the objective function (\ref{Fix_X_Y})  w.r.t.
%$\mathbf{B}_{k}$ and setting it to zero, the closed-form solution of $\mathbf{B}_{k}$ can
%be obtained as
\begin{equation}\label{solution_B_k}
		\begin{aligned}
			\begin{split}
\mathbf{B}_k^{\mathrm{opt}} &= \left(  2\sum_{m=1}^{K}\mathbf{H}_m^{\rm{H}}\mathbf{U}_m\mathbf{W}_m\mathbf{U}_m^{\rm{H}}\mathbf{H}_m  + \frac{1}{\rho^{(t)}} \left( \mathbf{I}_{N_{\rm{ t}}}  + \mathbf{V}^{\rm{H}}\mathbf{V}\right)  \right) ^{-1} \\ &\quad\times \left(\frac{1}{\rho^{(t)}}\left( \mathbf{X}_{k} + \mathbf{V}^{\rm{H}}\mathbf{Y}_k\right) + 2\mathbf{H}_{k}^{\rm{H}}\mathbf{U}_{k}\mathbf{W}_{k} \right).
			\end{split}
		\end{aligned}
\end{equation}
\paragraph{With fixed $\left\{ \mathbf{B}_{k}\right\}$, optimize $\left\{\mathbf{X}_k, \mathbf{Y}_k \right\}$}For given transmit beamforming matrices $\left\{ \mathbf{B}_{k}\right\}$, we can get the following optimization subproblem { in which} only items related to $\left\{\mathbf{X}_k\right\}$ and $\left\{\mathbf{Y}_k\right\}$ are kept 
\begin{subequations}\label{Fix_B_k}
	\begin{align}
		\mathop {\min }\limits_{\left\lbrace  {\mathbf{X}_k,\mathbf{Y}_k}\right\rbrace  }\quad 
		&\frac{1}{2\rho^{(t)}}\left( \sum_{k=1}^K\Vert {\mathbf{X}_k} - \mathbf{B}_{k}\Vert_{F}^{2}+ \sum_{k=1}^K\Vert {\mathbf{Y}_k} -  \mathbf{V}\mathbf{B}_{k}\Vert_{F}^{2}\right) 
		\\
		 \textrm{s.t.}\quad\quad
		&(\ref{X_Con}), (\ref{Y_Con}).
	\end{align}
\end{subequations}

Since optimization variables w.r.t. different blocks $\left\{\mathbf{X}_{k}\right\}$  and $\left\{\mathbf{Y}_k\right\}$ are separated in both the objective function and constraints. Therefore, Problem (\ref{Fix_B_k}) can be divided into two separated subproblems, which can be solved in a parallel manner as follows.
\begin{algorithm}
	\caption{Bisection search method to solve Problem (\ref{ADMM_Y_k}).}
	\label{alg-1}
	\begin{algorithmic}[1]
		\State {Initialize the accuracy $\eta,$ the bounds $\mu_{{\rm{ lb}}}$ and $\mu_{{\rm{ ub}}}$; }
		\State {If $h(0) \geq \sigma_\mathrm{r}^{2}\gamma_{0}$ holds, the auxiliary variable $\mathbf{Y}_k$ is given by $\mathbf{Y}_k^{\mathrm{opt}}=\mathbf{Y}_k\left( 0\right), \forall k$ and terminate; Othersize, go to Step 3;}
		\State {Calculate $\mu =  \left(\mu_{{\rm{ lb}}}+\mu_{{\rm{ ub}}} \right) / 2$;}
		\State {If $h({\mu})\geq \sigma_\mathrm{r}^{2}\gamma_{0}$, set ${\mu _{\rm{ub}}}={\mu}$. Otherwise, set ${\mu _{\rm{lb}}}={\mu}$; }
		\State {If  $\left| {\mu_{\rm{lb}} - \mu_{\rm{ub}}} \right| \le \eta $, terminate. Otherwise, go to Step 2.}
	\end{algorithmic}
\end{algorithm}

First, our focus turns to the optimization problem associated with $\left\{\mathbf{Y}_{k}\right\}$, which is expressed as follows:
%On the one hand, the subproblem regarding to block  $\left\{\mathbf{Y}_{k}\right\}$ is formulated by
\begin{subequations}\label{ADMM_Y_k}
	\begin{align}
		\mathop {\min }\limits_{\left\lbrace {\mathbf{Y}_k} \right\rbrace  }\quad 
		&\sum_{k=1}^K\Vert {\mathbf{Y}_k} -  \mathbf{V}\mathbf{B}_{k}\Vert_{F}^{2} 
		\\
		\quad\textrm{s.t.}\quad
		&(\ref{Y_Con}) \label{ADMM_Y_c}.
	\end{align}
\end{subequations}

It has been noted that Problem (\ref{ADMM_Y_k}) constitutes a quadratic constraint quadratic programming (QCQP) problem. To find the optimal solution for Problem (\ref{ADMM_Y_k}), one can solve its corresponding dual problem\cite{boyd2004convex}. We formulate the Lagrangian function of Problem (\ref{ADMM_Y_k})  associated with Lagrange multiplier  $\mu$ ($\mu\geq 0 $) as 
%It can be observed that Problem (\ref{ADMM_Y_k}) is  a QCQP problem with a quadratic objective and one quadratic inequality constraint. Fortunately, it was shown in \cite[Appendix B.1]{boyd2004convex} that the strong duality holds for any optimization problem with a  quadratic objective and one quadratic inequality constraint, provided Slater’s condition holds. The optimal solution to Problem (\ref{ADMM_Y_k}) can be obtained by solving its dual problem. By introducing dual variable $\mu\left(\mu \geq 0\right)$ associated with constraint (\ref{ADMM_Y_c}), the Lagrangian function of Problem (\ref{ADMM_Y_k}) is given by
%\begin{eqnarray}\label{L_2}
%	\mathcal{L}_2\left(\mathbf{Y}_{k}, \mu\right)=\sum_{k=1}^K\Vert {\mathbf{Y}_k} -  \mathbf{V}\mathbf{B}_{k}\Vert_{F}^{2} \nonumber\\  + \mu \left( \sigma_r^{2}\gamma_0  -\sum_{k=1}^{K}\operatorname{tr}\left(\mathbf{Y}_k\mathbf{Y}_k^{\rm{H}}\right)\right  ) 
%\end{eqnarray}
\begin{eqnarray}\label{L_2}
	\mathcal{L}_1\left(\mathbf{Y}_{k}, \mu\right)&=&\sum_{k=1}^K(1-\mu) \Vert {\mathbf{Y}_k}\Vert_{F}^{2}-\sum_{k=1}^K2\operatorname{Re}\left(  \mathbf{B}_{k}^{\rm{H}}\mathbf{V}^{\rm{H}}{\mathbf{Y}_k}\right)  \nonumber		
	\\&+& \sum_{k=1}^K \Vert {\mathbf{V}\mathbf{B}_{k}}\Vert_{F}^{2}  +\mu\sigma_{{\rm{ r}}}^2\gamma_0.
\end{eqnarray}

Define  $f_1\left(\mu\right)=\min _{\mathbf{Y}_{k}} \mathcal{L}_1\left(\mathbf{Y}_{k}, \mu\right)$ as the dual function of Problem (\ref{ADMM_Y_k}). Besides, to ensure the boundedness of the dual function $f_1(\mu)$, we can deduce that $0 \leq \mu<1$. Once we set $\mu > 1$ meaning  $1-\mu$, which  is the coefficient in front of the $\Vert {\mathbf{Y}_k}\Vert_{F}^{2}$ less than zero. In that case, if we set  ${\mathbf{Y}_{k}} = \varrho \mathbf{I}$ and let $\varrho$  be positive infinity, the $f_1\left(\mu\right)$ is unbounded. Similar considerations apply for $\mu = 1.$ Accordingly, the Lagrange multiplier $\mu$ is constrained within the range of $0 \leq \mu < 1.$ 
%To make dual function  $f_1\left(\mu\right)=\min _{\mathbf{Y}_{k}} \mathcal{L}_1\left(\mathbf{Y}_{k}, \mu\right)$ bounded, we should make $1-\mu> 0$, i.e., $\mu <1$. Otherwise, we can always set ${\mathbf{Y}_{k}} = \kappa \mathbf{I}$ and let $\kappa$  to be positive infinity, which will make $f_1\left(\mu\right)$ unbounded. This thus completes the proof of Lemma 1. 

By leveraging the first-order optimality conditions, we can determine the best solution for $f_1\left(\mu\right)$ as
\begin{eqnarray}\label{Y_k_solution}
	\mathbf{Y}_k^{\mathrm{opt}}\left(\mu \right)  = \frac{\mathbf{V}\mathbf{B}_k}{1-\mu},\quad k=1,\cdots,K.
\end{eqnarray}

Then, we define 
\begin{eqnarray}\label{Y_h}
	h\left(\mu \right) & \triangleq & \sum_{k=1}^{K} \operatorname{tr}\left(	\mathbf{Y}_k\left(\mu \right)	\mathbf{Y}_k^{\rm{H}}\left(\mu \right) \right) \nonumber
	\\
	&=&\operatorname{tr}\left(\left(1-\mu \right)^{-2}  \mathbf{\Gamma}\right)\nonumber \\
	&=&\sum_{i=1}^{N_{\rm{ t}}}\frac{[\mathbf{\Gamma}]_{i,i}}{\left( 1-\mu\right) ^{2}},
\end{eqnarray}
where $\mathbf{\Gamma} = \sum_{k=1}^{K}\mathbf{V}\mathbf{B}_k\mathbf{B}_k^{\rm{H}}\mathbf{V}^{\rm{H}}$ and $[\mathbf{\Gamma}]_{i,i}$ denotes the $i$-th diagonal element of matrix $\mathbf{\Gamma}$. It is easy to confirm that $h\left(\mu\right)$ exhibits a strictly increasing trend for $0\leq\mu <1$. If $  h\left( 0\right) > \sigma^{2}_\mathrm{r}\gamma_0$, then the corresponding optimal dual variable  $\mu^{\mathrm{opt}} = 0$. Otherwise, according to  the complementary condition of $\mu \left( \sigma_\mathrm{r}^{2}\gamma_0  - h\left(\mu \right)\right) = 0,$ we need to find a positive  $\mu$ according to the following
equation:
\begin{eqnarray}\label{h_eq}
	h\left(\mu \right)=\sigma_\mathrm{r}^{2}\gamma_{0}.
\end{eqnarray}

By exploiting the monotonic property of $h\left( \mu\right) $, the solution
$  \mu^{\mathrm{opt}} $  can be easily obtained using a straightforward bisection search method  \cite{pan2020multicell}, ranging from $0$ to $1$. 
%\begin{eqnarray}\label{S_mu}
%\mu \left( \sigma_r^{2}\gamma_0  -\sum_{k=1}^{K} \operatorname{tr}\left(\mathbf{Y}_k\mathbf{Y}_k^{\rm{H}}\right)\right) = 0.
%\end{eqnarray}
The overall algorithm to solve Problem (\ref{ADMM_Y_k}) is summarized
in Algorithm  \ref{alg-1}.

%On the other hand, the subproblem regarding to the block $\left\{\mathbf{X}_k\right\}$ is given by
%\begin{subequations}\label{ADMM_X_k}
%	\begin{align}
%			\mathop {\min }\limits_{\left\lbrace {\mathbf{X}_k} \right\rbrace }\quad 
%			&\sum_{k=1}^K\Vert {\mathbf{X}_k} - \mathbf{B}_{k}\Vert_{F}^{2}
%			\\
%			\quad\textrm{s.t.}\quad
%			&\sum_{k=1}^K \operatorname{tr}\left(\mathbf{X}_k\mathbf{X}_k^{\rm{H}}\right) \leq P_0\label{ADMM_X_c}
%		\end{align}
%\end{subequations}
%The optimal solution to Problem (\ref{ADMM_X_k}) can be obtained by solving its dual problem which is  similar to  Algorithm \ref{alg-1} and is omitted here for brevity.

Second, the subproblem associated with $\left\{\mathbf{X}_k\right\}$ is given by
\begin{subequations}\label{ADMM_X_k}
	\begin{align}
				\mathop {\min }\limits_ {\left\lbrace \mathbf{X}_k\right\rbrace }\quad 
			&\sum_{k=1}^K\Vert {\mathbf{X}_k} - \mathbf{B}_{k}\Vert_{F}^{2}
			\\
			\quad\textrm{s.t.}\quad
			&(\ref{X_Con}).\label{ADMM_X_c}
		\end{align}
\end{subequations}

Fortunately, Problem (\ref{ADMM_X_k}) is a convex problem which can be solved by solving its dual problem. We formulate the Lagrangian function of Problem (\ref{ADMM_X_k})  associated with Lagrange multiplier $\tau\left(\tau \geq 0\right)$ as 
\begin{equation}\label{L1}
\mathcal{L}_2\left(\mathbf{X}_{k}, \tau \right)=\sum_{k=1}^{K}\Vert {\mathbf{X}_k} - \mathbf{B}_{k}\Vert^{2}_{F} +\tau\left(\sum_{k=1}^K \operatorname{tr}\left(\mathbf{X}_k\mathbf{X}_k^{\rm{H}}\right) - P_0\right).
\end{equation}

Define  $f_2\left(\tau\right)=\min _{\mathbf{X}_{k}} \mathcal{L}_2\left(\mathbf{X}_{k}, \tau\right)$ as the dual function of Problem (\ref{ADMM_X_k}).

By leveraging the first-order optimality conditions, we can determine the best solution for $f_2\left(\tau\right)$ as
%The optimal solution to $f_2\left(\lambda\right)$ can be obtained by leveraging the first-order optimality conditions. Specifically, by taking the first-order derivative of $\mathcal{L}_2\left(\mathbf{X}_{k}, \lambda\right)$ w.r.t. $\mathbf{X}_{k}$ and setting it to zero, we obtain the optimal solution as
\begin{eqnarray}\label{X_k_solution}
	\mathbf{X}_k^{{\rm{ opt}}}\left(\tau \right)  = \frac{\mathbf{B}_k}{1+\tau},\quad k = 1\cdots,K.
\end{eqnarray}

Similarly, we define
\begin{eqnarray}\label{X_g}
	g\left(\tau \right) & \triangleq & \sum_{k=1}^{K} \operatorname{tr}\left(\mathbf{X}_k\left(\tau \right) \mathbf{X}_k^{\rm{H}}\left(\tau \right) \right) \nonumber
	\\
	&=&\operatorname{tr}\left(\left(1+\tau\right)^{-2}  \mathbf{M}\right) \\
	&=&\sum_{i=1}^{D_k}\frac{[\mathbf{M}]_{i,i}}{\left( 1+\tau\right) ^{2}},
\end{eqnarray}
where $\mathbf{M} = \sum_{k=1}^{K}\mathbf{B}_k^{\rm{H}}\mathbf{B}_k$ and $[\mathbf{M}]_{i,i}$ denote the $i$-th diagonal element of matrix $\mathbf{M}$. It can be easily confirmed that $g(\tau)$ exhibits a monotonically decreasing tend for  $\tau \geq 0$. Hence, if $g(0) \leq P_{0}$, the corresponding optimal dual variable  $\tau^{\mathrm{opt}} = 0$. Otherwise, according to the complementary condition: 
\begin{eqnarray}\label{S_lambda}
	\tau\left(\sum_{k=1}^K \operatorname{tr}\left(\mathbf{X}_k\left(\tau \right) \mathbf{X}_k^{\rm{H}}\left(\tau \right) \right) - P_0\right) = 0,
\end{eqnarray}
we need to find a positive  $\tau$ by using the bisection based search method according to the following equation:
\begin{eqnarray}
	g\left(\tau \right)=\sum_{i=1}^{D_k}\frac{[\mathbf{M}]_{i,i}}{\left( 1+\tau_{{\rm{ ub}}}\right) ^{2}}  = P_0.
\end{eqnarray}

%Otherwise, the optimal $\lambda$ can be obtained by using the bisection based search method to find the solution of the following equation:
%
%The optimal dual variable $\lambda$ should be chosen for ensuring that the following complementary slackness condition is satisfied:
%\begin{eqnarray}\label{S_lambda}
%\lambda\left(\sum_{k=1}^K \operatorname{tr}\left(\mathbf{X}_k\left(\lambda \right) \mathbf{X}_k\left(\lambda \right)^{\rm{H}} \right) - P_0\right) = 0.
%\end{eqnarray}

To apply the bisection search method, we have to find the upper bound of $\tau$, which is given by
\begin{eqnarray}\label{lambda_up}
\tau <\sqrt{\frac{\sum_{i=1}^{D_k}[\mathbf{M}]_{i,i}}{P_0}}  \triangleq \tau_{{\rm{ ub}}}.
\end{eqnarray}
This can be proved as follows:
\begin{eqnarray}\label{prove_lambda_up}
	g\left(\tau_{{\rm{ ub}}} \right)=\sum_{i=1}^{D_k}\frac{[\mathbf{M}]_{i,i}}{\left( 1+\tau_{{\rm{ ub}}}\right) ^{2}} <\frac{\sum_{i=1}^{D_k}[\mathbf{M}]_{i,i}}{\tau_{{\rm{ ub}}}^{2} } = P_0.
\end{eqnarray}

\begin{algorithm}
	\caption{Penalty-based algorithm for solving Problem (\ref{otp_bk}).}\label{alg-2}
	\begin{algorithmic}[1]
		\State {Initialize $\left\lbrace  \mathbf{B}_k^{\left[ 0 \right] } \right\rbrace = \left\lbrace \mathbf{B}_k^{\left( \mathrm{n}\right) }\right\rbrace$ ,$\left\lbrace  \mathbf{X}_k^{\left[  0\right]  } \right\rbrace = \left\lbrace \mathbf{B}_k^{\left( \mathrm{n}\right) }\right\rbrace$, $
		\left\lbrace \mathbf{Y}_k^{\left[ 0\right]  } \right\rbrace=\left\lbrace  \mathbf{V}\mathbf{B}_{k}^{\left( \mathrm{n}\right) }\right\rbrace$, $t=0$, the penalty coefficient $\rho^{0}$, step size $c$ , tolerance of accuracy $\xi$ and $\epsilon$.}
		\While{the fractional increase of (\ref{B_obj}) is larger than $\epsilon$ or penalty terms larger than $\xi$}
		\State {Update  $\mathbf{B}_k^{[t+1]}$ by using (\ref{solution_B_k});}
		\State {Update   $\mathbf{Y}_k^{[t+1]}$ via Algorithm \ref{alg-1}; }
		\State {Update  $\mathbf{X}_k^{[t+1]}$ by solving Problem (\ref{ADMM_X_k});}
		\State {Set $\rho^{[t+1]}=c \rho^{[t]}, t \leftarrow t+1 $.}
		\EndWhile
		\Ensure $\left\lbrace  \mathbf{B}_k^{ \left( n+1 \right)  } \right\rbrace = \left\lbrace \mathbf{B}_k^{ \left[ t\right]  }\right\rbrace $.
	\end{algorithmic}
\end{algorithm}

%Repeat
%\Repeat
%\State {Update transmit beamformers $\mathbf{B}_k^{[t+1]}$ based on (\ref{solution_B_k});}
%\State {Update  auxiliary variable $\mathbf{Y}_k^{[t+1]}$ by solving Problem (\ref{ADMM_Y_k}) through  Algorithm \ref{alg-1}; }
%\State {Update  auxiliary variable $\mathbf{X}_k^{[t+1]}$ by solving Problem (\ref{ADMM_X_k});}
%\Until{ the fractional decrease of the objective value of (\ref{Problem_with_auxi_in_obj}) is below the threshold $ \epsilon$.}
%\State {Update $\rho^{[t+1]}=c \rho^{[t]}, t \leftarrow t+1 $.}
%\Until{termination indicator is below a  threshold
%	$\xi$.}
\subsection{Optimize  the Phase Shift  Matrix $\bf{\Theta}$}
In the following, our primary focus lies in optimizing matrix $\boldsymbol{\Theta}$ while keeping $\left\lbrace {\bf{B}}_k\right\rbrace $, $\left\lbrace {\mathbf{W}}_k\right\rbrace $, and $\left\lbrace {\mathbf{U}}_k\right\rbrace $ fixed. We first substitute $\mathbf{E}_k$ into (\ref{sum_MSE}) and eliminate the constant terms, we can formulate the optimization problem for phase shift as follows:
%In this subsection, we focus our attention on optimizing the  phase shifts $\boldsymbol{\Theta}$, while fixing ${\bf{B}}_k$, $ {\mathbf{W}}$, and $ {\mathbf{U}}_k $. By substituting $\mathbf{E_k}$ into (\ref{sum_MSE}) and removing the constant terms, the phase shift optimization problem is formulated as:
\begin{subequations}\label{otp-theta}
	\begin{align}
		\mathop {\min }\limits_{{\boldsymbol{\Theta}}}\quad &-\sum_{k=1}^K\operatorname{tr}\left(\mathbf{W}_k\mathbf{U}_k^{\rm{H}}\mathbf{H}_k\mathbf{B}_k\right)\nonumber
		\\
		&+\sum_{k=1}^K\operatorname{tr}\left(\mathbf{W}_k\mathbf{U}_k^{\rm{H}}\mathbf{H}_k\sum\limits_{i = 1}^K  {{\bf{B}}_{i}}{\bf{B}}_{i}^{\rm{H}}\mathbf{H}_k^{\rm{H}}\mathbf{U}_k\right)\nonumber
		\\
		&-\sum_{k=1}^K\operatorname{tr}\left(\mathbf{W}_k\mathbf{B}_k^{\rm{H}}\mathbf{H}_k^{\rm{H}}\mathbf{U}_k\right)\label{theta_obj}
		\\
		\qquad\ \textrm{s.t.}\quad
		&(\ref{SNR}),(\ref{phase}).
	\end{align}
\end{subequations}

First, let us define ${\bf{B}} = \sum_{k=1}^K{\bf{B}}_{k}{\bf{B}}_{k}^{\rm{H}}$, by substituting ${\bf{H}}_{k} =   {\bf{H}}_{{\rm{ bu}},k} + {\bf{H}}_{{\rm{ ru}},k} {\mathbf{\Theta}}^{\rm{H}} {\bf{G}}  $ into 
$\mathbf{W}_k\mathbf{U}_k^{\rm{H}}\mathbf{H}_k\sum\limits_{i = 1}^K  {{\bf{B}}_{i}}{\bf{B}}_{i}^{\rm{H}}\mathbf{H}_k^{\rm{H}}\mathbf{U}_k$, we have
\begin{equation}
	\begin{aligned}\label{theta_long_part}
	\mathbf{W}_k\mathbf{U}_k^{\rm{H}}\mathbf{H}_k\sum\limits_{i = 1}^K  {{\bf{B}}_{i}}&{\bf{B}}_{i}^{\rm{H}}\mathbf{H}_k^{\rm{H}}\mathbf{U}_k
	=\mathbf{W}_k\mathbf{U}_k^{\rm{H}}\mathbf{H}_{{\rm{ bu}},k}{{\bf{B}}}\mathbf{H}_{{\rm{ bu}},k}^{\rm{H}}\mathbf{U}_k
	\\
	\qquad\qquad\qquad+&\mathbf{W}_k \mathbf{U}_k^{\mathrm{H}} \mathbf{H}_{{\rm{ r u}}, k} \mathbf{\Theta}^{\mathrm{H}} \mathbf{G} \mathbf{B G}^{\mathrm{H}} \mathbf{\Theta} \mathbf{H}_{{\rm{ r u}}, k}^{\mathrm{H}} \mathbf{U}_k
	\\
	+&\mathbf{W}_k\mathbf{U}_k^{\rm{H}}\mathbf{H}_{{\rm{ ru}},k}{\mathbf{\Theta}}^{\rm{H}}{{\bf{G}}} {{\bf{B}}}\mathbf{H}_{{\rm{ bu}},k}^{\rm{H}}\mathbf{U}_k \qquad
	\\
	+&\mathbf{W}_k\mathbf{U}_k^{\rm{H}}\mathbf{H}_{{\rm{ bu}},k}{{\bf{B}}}{{\bf{G}}}^{\rm{H}} {\mathbf{\Theta}}\mathbf{H}_{{\rm{ ru}},k}^{\rm{H}}\mathbf{U}_k\qquad\qquad\qquad
	\end{aligned}
\end{equation}
and 
\begin{align}\label{theta_short_part}
	%\begin{aligned}
	\mathbf{W}_k\mathbf{U}_k^{\rm{H}}\mathbf{H}_k\mathbf{B}_k=\mathbf{W}_k\mathbf{U}_k^{\rm{H}}\mathbf{H}_{{\rm{ ru}},k}{\mathbf{\Theta}}^{\rm{H}}{{\bf{G}}}\mathbf{B}_k+\mathbf{W}_k\mathbf{U}_k^{\rm{H}}\mathbf{H}_{{\rm{ bu}},k}\mathbf{B}_k.
	%\end{aligned}
\end{align}

Then, by defining $\mathbf{C}_{k} = \mathbf{H}_{{\rm{ ru}},k}^{\rm{H}}\mathbf{U}_k	\mathbf{W}_k\mathbf{U}_k^{\rm{H}}\mathbf{H}_{{\rm{ ru}},k}$, $\mathbf{D} = {{\bf{G}}} {{\bf{B}}}{{\bf{G}}}^{\rm{H}}$ and $\mathbf{E}_{k}^{\rm{H}} = {{\bf{G}}} {{\bf{B}}}\mathbf{H}_{{\rm{ bu}},k}^{\rm{H}}\mathbf{U}_k\mathbf{W}_k\mathbf{U}_k^{\rm{H}}\mathbf{H}_{{\rm{ ru}},k} $, and using (\ref{theta_long_part}), we can obtain
\begin{equation}
\begin{aligned}\label{theta_long_complete}
\operatorname{tr}( \mathbf{W}_k\mathbf{U}_k^{\rm{H}}\mathbf{H}_k\sum\limits_{i = 1}^K  {{\bf{B}}_{i}}{\bf{B}}_{i}^{\rm{H}}\mathbf{H}_k^{\rm{H}}\mathbf{U}_k) =
\operatorname{tr}\left(\mathbf{C}_{k} {\mathbf{\Theta}}^{\rm{H}} \mathbf{D}  {\mathbf{\Theta}} \right) + \operatorname{tr}\left( \mathbf{E}_{k}{\mathbf{\Theta}}\right)
\\ +
\operatorname{tr}\left(\mathbf{E}_{k}^{\rm{H}}{\mathbf{\Theta}}^{\rm{H}}\right) + \underbrace{\operatorname{tr}(\mathbf{W}_k\mathbf{U}_k^{\rm{H}}\mathbf{H}_{{\rm{ bu}},k}{\sum_{k=i}^K{\bf{B}}_{i}{\bf{B}}_{i}^{\rm{H}}}\mathbf{H}_{{\rm{ bu}},k}^{\rm{H}}\mathbf{U}_k)}_{\rm{const}_{1}},
	\end{aligned}
\end{equation}
where ${\rm{const}_{1}}$ is a constant term that does not depend on $\mathbf{\Theta}$.

Similarly, define $\mathbf{F}_{k}^{\rm{H}} = \mathbf{G}\mathbf{B}_k	\mathbf{W}_{k}\mathbf{U}_k^{\rm{H}}\mathbf{H}_{{\rm{ ru}},k}$,  we can obtain
\begin{equation}\label{theta_short_complete}
\operatorname{tr}\left(\mathbf{W}_k\mathbf{U}_k^{\rm{H}}\mathbf{H}_k\mathbf{B}_k\right) = \operatorname{tr}\left(\mathbf{\Theta}^{\rm{H}}\mathbf{F}_{k}^{\rm{H}}\right) + \underbrace{\operatorname{tr}\left(\mathbf{W}_k\mathbf{U}_k^{\rm{H}}\mathbf{H}_{{\rm{ bu}},k}\mathbf{B}_k\right)}_{\rm{const}_{2}},
\end{equation}
where ${\rm{const}_{2}}$ is a  constant term that does not depend on $\mathbf{\Theta}$.

From  (\ref{theta_long_complete}) and (\ref{theta_short_complete}), the objective function (\ref{theta_obj}) can be derived as:
\begin{equation}\label{theta_all_complete}
	\operatorname{tr}\left(\mathbf{C}\mathbf{\Theta}^{\rm{H}}\mathbf{D}\mathbf{\Theta}\right) +\operatorname{tr}\left(\mathbf{Z}\mathbf{\Theta}^{\rm{H}}\right)+\operatorname{tr}\left(\mathbf{Z}^{\rm{H}}\mathbf{\Theta}\right),
\end{equation}
where $\mathbf{C} = \sum_{k=1}^{K}\mathbf{C}_{k}$ and $\mathbf{Z} = \sum_{k=1}^{K}\left(\mathbf{E}_{k}^{\rm{H}}-\mathbf{F}_{k}^{\rm{H}}\right)$. Furthermore, using ${\boldsymbol{\vartheta}} $, we arrive at \cite[Eq. (1.10.6)]{zhang2017matrix}
%We define  ${\boldsymbol{\vartheta}} =  \left\{ {{e^{j{\theta _1}}}, \cdots ,{e^{j{\theta _l}}}, \cdots ,{e^{j{\theta _L}}}} \right\}$ as the diagonal elements vector of $\mathbf{\Theta}$, and use the matrix identity of \cite[Eq. (1.10.6)]{zhang2017matrix}, we arrive at
\begin{equation}\label{theta_complete_first_trs}
	\operatorname{tr}\left( \mathbf{C}\mathbf{\Theta}^{\rm{H}}\mathbf{D}\mathbf{\Theta}\right) ={\boldsymbol{\vartheta}}^{\mathrm{H}}\mathbf{\Xi} {\boldsymbol{\vartheta}},
\end{equation}
where $\mathbf{\Xi} = \mathbf{D} \odot \mathbf{C}^{\mathrm{T}}$.

Let $\mathbf{z}$ be the collection of diagonal elements of matrix $\mathbf{Z}$, given by $\mathbf{z}=\operatorname{diag}{\left( \mathbf{Z}\right) }$. Then, we have
%\left[[\mathbf{Z}]_{1,1}, \cdots,[\mathbf{Z}]_{L, L}\right]^{\mathrm{T}}
\begin{eqnarray}\label{theta_complete_second_trs}
\operatorname{tr}(\mathbf{\Theta} \mathbf{Z}^{\rm{H}})=\mathbf{z}^{\mathrm{H}}{\boldsymbol{\vartheta}}.
\end{eqnarray}
% , \operatorname{tr}\left(\mathbf{\Theta}^{\mathrm{H}} \mathbf{Z}\right)={\boldsymbol{\vartheta}}^{\mathrm{H}}\mathbf{z}  

Hence, the objective function of Problem (\ref{otp-theta}) can be rewritten as
%\begin{eqnarray}\label{theta_sec-order_obj}
${\boldsymbol{\vartheta}}^{\mathrm{H}} \mathbf{\Xi}{\boldsymbol{\vartheta}}+\mathbf{z}^{\mathrm{H}}{\boldsymbol{\vartheta}}+{\boldsymbol{\vartheta}}^{\mathrm{H}}\mathbf{z}.$
%\end{eqnarray}

In the following, we reformulate the left-hand-side (LHS) of the radar SNR constraint (\ref{SNR}) in a more tractable way. According to the equation $\operatorname{tr}\left(\mathbf{ABCD}\right) = \left(\operatorname{vec}\left(\mathbf{D}^{\rm{T}}\right)\right)^{\rm{T}} (
\mathbf{C}^{\rm{T}}  \otimes 
\mathbf{A})\operatorname{vec}\left( {\mathbf{B}}\right) $,  we have
\begin{equation}\label{theta_c_gam_trs}
	\begin{aligned}
		\operatorname{tr}\left({\bf{V}}^{\rm{H}}{\bf{V}}\mathbf{B}_{k} \mathbf{B}_{k}^{\rm{H}}\right)    
		=&( \operatorname {vec} ( ( \operatorname { \mathbf{\Theta} \mathbf{A}\mathbf{\Theta}^{\rm{H}} }  )^ {\rm{ T }} )) ^ { \rm{T} }( (\mathbf{G}^{*}\mathbf{G}^{\rm{T}} )\\ &\otimes\left( \mathbf{G}\mathbf{B}_{k} \mathbf{B}_{k}^{\rm{H}}\mathbf{G}^{\rm{H}}\right) )  \operatorname{vec}\left(\mathbf{\Theta} \mathbf{A}\mathbf{\Theta}^{\rm{H}}\right).
	\end{aligned}
\end{equation}
%\begin{equation}\label{theta_c_gam_trs}
%	\begin{aligned}
%		&\operatorname{tr}\left({\bf{V}}^{\rm{H}}{\bf{V}}\mathbf{B}_{k} \mathbf{B}_{k}^{\rm{H}}\right)    \\
%		=& \operatorname{tr}\left(\mathbf{B}_{k}^{\rm{H}}\left(\mathbf{G}^{\mathrm{H}} \mathbf{\Theta A} \boldsymbol{\Theta}^{\mathrm{H}} \mathbf{G}\right) ^{\rm{H}} \left(\mathbf{G}^{\mathrm{H}} \mathbf{\Theta A} \boldsymbol{\Theta}^{\mathrm{H}} \mathbf{G}\right) \mathbf{B}_{k} \right) \\
%		=&( \operatorname {vec} ( ( \operatorname { \mathbf{\Theta} \mathbf{A}\mathbf{\Theta}^{\rm{H}} }  )^ {\rm{ T }} )) ^ { \rm{T} }\left( \left(\mathbf{G}^{*}\mathbf{G}^{\rm{T}} \right)  \otimes\left( \mathbf{G}\mathbf{B}_{k} \mathbf{B}_{k}^{\rm{H}}\mathbf{G}^{\rm{H}}\right) \right) \\&\operatorname{vec}\left(\mathbf{\Theta} \mathbf{A}\mathbf{\Theta}^{\rm{H}}  \right).
%	\end{aligned}
%\end{equation}

As $\mathbf{A}$ is a Hermitian matrix, after some further manipulations, we obtain 
\begin{equation}
	( \operatorname {vec} ((\operatorname { \mathbf{\Theta} \mathbf{A}\mathbf{\Theta}^{\rm{H}}} )^ {\rm{ T }} )) ^ { \rm{T} } = (\operatorname {vec} ( \operatorname {\mathbf{\Theta} \mathbf{A}\mathbf{\Theta}^{\rm{H}}}))  ^ { \rm{H} }=\operatorname{Diag}\left( {\operatorname{vec}\left( \mathbf{A}\right) }\right)\hat{\boldsymbol{\vartheta}}, 
\end{equation}
where   $\hat{\boldsymbol{\vartheta}} $ is a vector of diagonal elements of the matrix $\left( \mathbf{\Theta}^{*}\otimes\mathbf{\Theta} \right) $.
%Since A is a Hermitian matrix, then
%\begin{equation}\label{theta_A}
%	\begin{aligned}
%		( \operatorname {vec} ( ( \operatorname { \mathbf{\Theta} \mathbf{A}\mathbf{\Theta}^{\rm{H}} }  )^ {\rm{ T }} )) ^ { \rm{T} } = (\operatorname {vec} ( \operatorname {\mathbf{\Theta} \mathbf{A}\mathbf{\Theta}^{\rm{H}}  }))  ^ { \rm{H} }. \\
%	\end{aligned}
%\end{equation}
%
%
%According to the characteristics of the diagonal matrix, and $\operatorname{vec}\left(\mathbf{ABC}\right))= \left(\mathbf{C}^{\rm{T}}\otimes\mathbf{A}\right) \operatorname{vec}\left(\mathbf{B}\right)$, we have
%
%\begin{equation}\label{theta_A}
%	\begin{aligned}
%		\operatorname {vec}\left( \mathbf{\Theta} \mathbf{A}\mathbf{\Theta}^{\rm{H}} \right) =\left(\mathbf{\Theta}^{*}\otimes\mathbf{\Theta}  \right) \operatorname{vec}\left(\mathbf{A} \right)  \\
%		=\operatorname{diag}\left( {\operatorname{vec}\left( \mathbf{A}\right) }\right) \hat{\boldsymbol{\vartheta}},
%	\end{aligned}
%\end{equation}
%where $ \hat{\boldsymbol{\vartheta}} $ is a vector of diagonal elements of the matrix $\left( \mathbf{\Theta}^{*}\otimes\mathbf{\Theta} \right) $. 
Thus, (\ref{theta_c_gam_trs}) can be expressed as
 \begin{equation}\label{theta_hat_get}
\hat{\boldsymbol{\vartheta}}^{\rm{H}}\mathbf{Q}_{k}\hat{\boldsymbol{\vartheta}} ,
\end{equation}
where
%\begin{equations}
	\begin{align}\label{theta_Q} 
	\mathbf{Q}_{k} &= 
	\left( \operatorname{Diag}\left( {\operatorname{vec}{\left( \mathbf{A}\right) }}\right) \right)^{\rm{H}} ( (\mathbf{G}^{*}\mathbf{G}^{\rm{T}} ) \nonumber\\
	 &\quad\otimes( \mathbf{G}\mathbf{B}_{k} \mathbf{B}_{k}^{\rm{H}}\mathbf{G}^{\rm{H}}) )
	\operatorname{Diag}( {\operatorname{vec}{( \mathbf{A}) }}) .
\end{align}
%\end{equations}

{Then, Problem } (\ref{otp-theta}) can be reformulated as
\begin{subequations}\label{re-opt-theta}
	\begin{align}
	\min _{\boldsymbol{\vartheta}}\quad&{\boldsymbol{\vartheta}}^{\mathrm{H}} \mathbf{\Xi} {\boldsymbol{\vartheta}}+\mathbf{z}^{\mathrm{H}}{\boldsymbol{\vartheta}}+{\boldsymbol{\vartheta}}^{\mathrm{H}}\mathbf{z}
		\\
		\qquad\ \textrm{s.t.}\quad
		&\sum_{k=1}^{K} \hat{\boldsymbol{\vartheta}} ^{\rm{H}}\mathbf{Q}_{k}\hat{\boldsymbol{\vartheta}}\geq \sigma_{\rm{ r}}^{2}\gamma_{0}\label{theta_re_c_gam},\\
%		& 0 \le {\theta _m} \le 2\pi, m = 1, \cdots ,L.
		& \left| {{\vartheta} _m} \right| = 1,  m\in \mathcal{L}.\label{unity}
	\end{align}
\end{subequations}

Solving Problem (\ref{re-opt-theta}) is challenging due to the presence of non-convex constraints (\ref{theta_re_c_gam}) and the unity-norm constraints (\ref{unity}). To overcome this challenge, we can employ the MM algorithm to address this issue. The solution procedure of the MM algorithm encompasses two steps\cite{sun2016majorization}.
%Problem (\ref{re-opt-theta}) is not easy to solve due to the non-convex  constraints (\ref{theta_re_c_gam})  and  the unity-norm constraints  (\ref{unity}). In order to circumvent this issue, the MM algorithm can be exploited to tackle this problem. 
%Solution process of the MM algorithm consists of two steps\cite{sun2016majorization}.
First, we find a surrogate function that locally approximates the constraint function (\ref{theta_re_c_gam}). In the second step, we optimize this optimization problem by replacing the constraint (\ref{theta_re_c_gam}) with the MM algorithm. We assume that $g(\mathbf{x})$ is an original constraint that needs to be dealt with, then the lower bound surrogate function $\overline{g}(\mathbf{x}|\mathbf{x}^{(t)})$ of the constraint should satisfy the following conditions in the $t$-th iteration:
\begin{enumerate}
	\item [1)] 
	$\overline{g}(\mathbf{x}^{(t)}|\mathbf{x}^{(t)}) = g(\mathbf{x}^{(t)}) $,
	\item  [2)] 
	$\nabla_\mathbf{x}\overline{g}(\mathbf{x}|\mathbf{x}^{(t)})|_{\mathbf{x}=\mathbf{x}^{(t)}} = \nabla_\mathbf{x}g(\mathbf{x})|_{\mathbf{x}=\mathbf{x}^{(t)}} $,
	\item  [3)]  
	$g(\mathbf{x}) \geq \overline{g}(\mathbf{x}|\mathbf{x}^{(t)})$.
	\end{enumerate}
%because the  variable in the objective function is ${\bf\vartheta}$ , which is the diagonal elements vector of $\mathbf{\Theta}$ ,but the variable in constraint (\ref{theta_re_c_gam}) is $\hat{\bf{\vartheta}} $, which is a vector of diagonal elements of the matrix $\left( \mathbf{\Theta}^{*}\otimes\mathbf{\Theta} \right) $.

Based on the above surrogate function conditions, we can use the  first-order Taylor expansion \cite{zhang2017matrix} to find out the surrogate function of  (\ref{theta_re_c_gam}). Accordingly, the surrogate function of  (\ref{theta_re_c_gam}) is given by
\begin{equation}\label{49}
	\begin{aligned}
		\sum_{k=1}^{K} \hat{\boldsymbol{\vartheta}}  ^{\rm{H}}\mathbf{Q}_{k}\hat{\boldsymbol{\vartheta}}&\geq \sum_{k=1}^{K}\left(2\operatorname{Re}{\left( \overline{\mathbf{q}}_{k}^{\rm{H}} \hat{\boldsymbol{\vartheta}}\right) } + C_{k} \right) \\
		&= \sum_{k=1}^{K}\left(2\operatorname{Re}\left( {\boldsymbol{\vartheta}}^\mathrm{H} \mathbf{M}_k {\boldsymbol{\vartheta}}\right) + C_{k}\right)
		\geq  \sigma_\mathrm{r}^{2}\gamma_{0},
	\end{aligned}
\end{equation}
where $\overline{\mathbf{q}}_{k}^{\rm{H}}  = \hat{\boldsymbol{\vartheta}}_t^{\rm{H}} {\mathbf{Q}}_{k}, C_{k} =  -\hat{\boldsymbol{\vartheta}}_t^{\rm{T}} {\mathbf{Q}}_{k}^{\rm{T}} \hat{\boldsymbol{\vartheta}}_{t}^{*}$ and $\mathbf{M}_{k}  = \operatorname{unvec}{\left(\overline{\mathbf{q}}_{k}^{\rm{H}} \right)}$. 
Problem  (\ref{re-opt-theta}) can be formulated as
%By substituting (\ref{50}) into (\ref{49}),
%the optimization problem in (\ref{re-opt-theta}) can be formulated as
\begin{subequations}\label{theta_after_all}
	\begin{align}
		\min _{\boldsymbol{\vartheta}}\quad&{\boldsymbol{\vartheta}}^{\mathrm{H}}\mathbf{\Xi} {\boldsymbol{\vartheta}}+\mathbf{z}^{\mathrm{H}}{\boldsymbol{\vartheta}}+{\boldsymbol{\vartheta}}^{\mathrm{H}}\mathbf{z}\label{theta_after_all_a}
		\\
		\qquad\ \textrm{s.t.}\quad
		&\sum_{k=1}^{K}2\operatorname{Re}{\left( {\boldsymbol{\vartheta}}^\mathrm{H} \mathbf{M}_k {\boldsymbol{\vartheta}}\right) } \geq \hat{\gamma}\label{51b},\\
		&\left| {{\vartheta} _m} \right| = 1,  m\in \mathcal{L},
	\end{align}
\end{subequations}
where $\hat{\gamma} = \sigma_\mathrm{r}^{2}\gamma_0 - \sum_{k=1}^{K}{C_{k}}$. Nevertheless, the unity-norm constraint in Problem (\ref{theta_after_all}) is difficult to deal with. To address this non-convex problem, we opt to convert it into an SDP problem and subsequently relax it to an SDR problem. First, we define $\mathbf{R}=\left[\begin{array}{cc}
	\mathbf{\Xi} &  \mathbf{z}^{}\\
	\mathbf{z}^{\rm{H}} &0
\end{array}\right]$, $\widetilde{\boldsymbol{\vartheta}} = \left[\begin{array}{l}
	\boldsymbol{\vartheta}\\
	h
\end{array}\right]$ and $ \mathbf{T}_{k} = \left[\begin{array}{cc}
	\mathbf{M}_{k}  & 0\\
	0&0
\end{array}\right]$.
Then introduce an auxiliary variable $h$ to transform  Problem (\ref{theta_after_all}) to a more tractable form as
%However, due to the unity-norm constrain, the problem in (\ref{theta_after_all}) is difficult to solve. To deal with this non-convex problem, we transform the aforementioned problem to the SDP problem and relax it to SDR problem.  For further processing, we introduce an auxiliary variable $t$ to transform  Problem (\ref{theta_after_all}) to a more tractable form. After some
%mathematical manipulations, the Problem (\ref{theta_after_all}) can be rewritten
%as 
\begin{subequations}\label{theta_SDP}
	\begin{align}
		\min _{\boldsymbol{\widetilde{\vartheta}}}\quad& \widetilde{\boldsymbol{\vartheta}}^{\rm{H}}\mathbf{R}\widetilde{\boldsymbol{\vartheta}}
		\\
		\qquad\ \textrm{s.t.}\quad
		&\sum_{k=1}^{K}2 \operatorname{Re}{\left(  \widetilde{\boldsymbol{\vartheta}}^{\rm{H}}\mathbf{T}_{k} \widetilde{\boldsymbol{\vartheta}}\right)  }\geq \hat{\gamma},\\
		&\left|  {{\widetilde{\vartheta}}}_m \right|= 1  , \forall m = 1, \cdots ,L+1,
	\end{align}
\end{subequations}
where the auxiliary variable $h$ satisfies $h^2=1$.

Subsequently, the problem in (\ref{theta_SDP}) is rewritten by dropping the rank-one constraint as
\begin{subequations}\label{theta_SDR}
	\begin{align}
		\min _{\boldsymbol{\widetilde{\mathbf{\Phi}}}}\quad& \operatorname{tr}\left( \widetilde{\mathbf{\Phi}}\mathbf{R}\right) 
		\\
		\qquad\ \textrm{s.t.}\quad
		&\sum_{k=1}^{K} 2\operatorname{Re}{\left( \operatorname{tr}\left( \widetilde{\mathbf{\Phi}}\mathbf{T}_{k}\right)\right) }\geq \hat{\gamma}\label{53b},\\
		&\widetilde{\mathbf{\Phi}}_{m,m}  = 1  ,\forall m= 1,\cdots,L+1,\\
		&\widetilde{\mathbf{\Phi}}\succeq0,
	\end{align}
\end{subequations}
where $\widetilde{\mathbf{\Phi}}= \widetilde{\boldsymbol{\vartheta}} \widetilde{\boldsymbol{\vartheta}}^{\rm{H}}$. Problem (\ref{theta_SDR}) is a convex problem that can be readily solved using the interior point method\cite{bomze2010interior}. After obtaining $\widetilde{\mathbf{\Phi}}$, we first
obtain the eigenvalue decomposition of $\widetilde{\mathbf{\Phi}}$ as { $\widetilde{\mathbf{\Phi}}=\mathbf{U} \mathbf{\Sigma} \mathbf{U}^{\rm{ H}}$, where $\mathbf{U}=\left[\mathbf{e}_1, \cdots, \mathbf{e}_{L+1}\right]$ }and $\boldsymbol{\Sigma}=\operatorname{Diag}\left(\left[ \lambda_1, \cdots, \lambda_{L+1}\right] \right)$ are a unitary matrix and a diagonal matrix, respectively, both with the size of $(L+1) \times(L+1)$. $\mathbf{e}$ and $\lambda$ are the eigenvectors and eigenvalues corresponding to $\widetilde{\mathbf{\Phi}}$. Subsequently, we derive a suboptimal solution for (\ref{theta_SDR}), represented as $\bar{\boldsymbol{\vartheta}} = \mathbf{U} \mathbf{\Sigma}^{1/2} \mathbf{r}$, where $\mathbf{r} \in \mathbb{C}^{(L+1) \times 1}$ denotes a random vector, satisfying $\mathcal{CN}\left(\mathbf{0}, \mathbf{I}_{L+1}\right)$.
%Then, we obtain a suboptimal solution to (\ref{theta_SDR}) as $\bar{\boldsymbol{\vartheta}}=\mathbf{U} \mathbf{\Sigma}^{1 / 2} \mathbf{r}$, where $\mathbf{r} \in \mathbb{C}^{(L+1) \times 1}$ is a random vector generated following the distribution of $\mathbf{r} \in \mathcal{C N}\left(\mathbf{0}, \boldsymbol{I}_{L+1}\right)$ and accordingly construct a series of candidate solutions as $\boldsymbol{\vartheta} = e^{j \operatorname{arg}([\frac{\bar{\boldsymbol{\vartheta}}}{\bar{\boldsymbol{\vartheta}}_{L+1}}](1:L))}$. 
Thereafter, we proceed to construct a series of candidate solutions, expressed as $\boldsymbol{\vartheta} = e^{j \operatorname{arg}([\frac{\bar{\boldsymbol{\vartheta}}}{\bar{{\vartheta}}_{L+1}}]_{(1:L)})}$. By independently generating Gaussian random vector  $\mathbf{r}$ multiple times, we obtain the solution of Problem (\ref{theta_after_all}) as the one achieving the minimum objective value of (\ref{theta_after_all_a}) while satisfying the SNR constraints (\ref{51b}) among all these random realizations.

The overall algorithm for solving Problem (\ref{comm_cen}) is summarized in Algorithm \ref{al_overall}.
\begin{algorithm}
	\caption{Overall algorithm for solving Problem (\ref{comm_cen}).}\label{al_overall}
	\begin{algorithmic}[1]
		\State {Initialize the iteration number $n= 1$, the maximum number of iterations $n_{\mathrm{max}},$ feasible $\mathbf{B}_k^{\left( 1\right) },\mathbf{\Theta}^{\left( 1\right) },$ error tolerance $\zeta$;}
		\Repeat
		\State {Given $\mathbf{B}_k^{\left( \mathrm{n}\right) } $ and $\mathbf{\Theta}^{\left( \mathrm{n}\right) },$ calculate the decoding matrices  $ \mathbf{U}_k^{\left( \mathrm{n}\right) } $ according to (\ref{decode_U}); }
		\State {Given $\mathbf{B}_k^{\left( \mathrm{n}\right) },  \mathbf{U}_k^{\left( \mathrm{n}\right) } $ and $\mathbf{\Theta}^{\left( \mathrm{n}\right) },$ calculate the auxiliary  matrices $ \mathbf{W}_k^{\left( \mathrm{n}\right) }$ according to (\ref{Weight_W}) ;}
		\State {Given $  {\mathbf{U}_k^{\left( \mathrm{n}\right)}},\mathbf{W}_k^{\left(\mathrm{ n}\right)} $ and $\mathbf{\Theta}^{\left( \mathrm{n}\right) },$ calculate the beamforming    matrices $ \mathbf{B}_k^{\left( \mathrm{n+1}\right) }$ through Algorithm \ref{alg-2};}
		\State {Given $  {\mathbf{U}_k^{\left( \mathrm{n}\right)}},\mathbf{W}_k^{\left( \mathrm{n}\right)} $ and $\mathbf{B}_k^{\left( \mathrm{n+1}\right) },$ calculate the optimal  $\mathbf{\Theta}^{\left(\mathrm{ n+1}\right) } $ by solving Problem (\ref{otp-theta});}
		\Until{{ $n > n_{\mathrm{max}}$ or $
				\left|E^{\left( n-1\right) } - E^{\left( n\right) }\right| < \zeta.\nonumber
				$ }}
	\end{algorithmic}
\end{algorithm}
%\begin{algorithm}
%	\caption{Overall algorithm for solving Problem (\ref{comm_cen}).}\label{al_overall}
%	\begin{algorithmic}[1]
%		\State {Initialize the iteration number $n= 1$, the maximum number of iterations $n_{max},$ feasible $\mathbf{B}_k^{\left( 1\right) },\mathbf{\Theta}^{\left( 1\right) },$ error tolerance $\zeta$;}
%		\Repeat
%		\State {Given $\mathbf{B}_k^{\left( n\right) } $ and $\mathbf{\Theta}^{\left( n\right) },$ calculate the decoding matrices  $ \mathbf{U}_k^{\left( n\right) } $ in (\ref{decode_U}); }
%		\State {Given $\mathbf{B}_k^{\left( n\right) },  \mathbf{U}_k^{\left( n\right) } $ and $\mathbf{\Theta}^{\left( n\right) },$ calculate the auxiliary  matrices $ \mathbf{W}_k^{\left( n\right) }$ ;}
%		\State {Given $  {\mathbf{U}_k^{\left( n\right)}},\mathbf{W}_k^{\left( n\right)} $ and $\mathbf{\Theta}^{\left( n\right) },$ calculate the beamforming    matrices $ \mathbf{B}_k^{\left( n+1\right) }$ by solving Problem (\ref{Problem_with_auxi_in_obj})  through  Algorithm \ref{alg-2};}
%		\State {Given $  {\mathbf{U}_k^{\left( n\right)}},\mathbf{W}_k^{\left( n\right)} $ and $\mathbf{B}_k^{\left( n+1\right) },$ calculate the optimal  $\mathbf{\Theta}^{\left( n+1\right) } $ by solving Problem (\ref{otp-theta})  ;}
%		\Until{If $n \geq n_{max}$ or \begin{eqnarray}
%				\left|E^{\left( n+1\right) } - E^{\left( n\right) }\right| \leq \zeta.\nonumber
%			\end{eqnarray}  terminate. Otherwise, set $ n \leftarrow n+1  $
%			and go to \textbf{Step} 2.}
%	\end{algorithmic}
%\end{algorithm}

Let us now analyze the complexity of the overall algorithm. In Step 3, the   complexity of computing the decoding matrices $ \mathbf{U}_k^{\left( n\right) } $ is  $o_1 = {\cal O}(KM_k^3)$. In Step 4, the complexity of calculating
the auxiliary matrices $ \mathbf{W}_k^{\left( n\right) }$ is given by $o_2= {\cal O}(KD_k^3)$. 

Then, the beamforming matrices $ \mathbf{B}_k^{\left( n+1\right) }$  is calculated through  Algorithm \ref{alg-2}. The penalty-based algorithm has double layers, where the maximum iteration numbers of the outer layer and inner layer are set to be  $I^o_{max}$ and $I_{max}$. Each iteration within the inner loop consists of three distinct steps. The complexity of calculating $\mathbf{B}_{k}$, $\mathbf{\Gamma}$ and $\mathbf{M}$  is ${\cal O}(KN_{\rm{ t}}^3)$, ${\cal O}(KN_{\rm{ t}}^2D_k)$ and ${\cal O}(KD_k^2N_{\rm{ t}})$  respectively.
%First, the complexity of computing the matrices $\mathbf{B}_{k}$ is ${\cal O}(KN_t^3)$. Second, for any pair of complex matrices $\mathbf{P} \in \mathbb{C}^{m \times n}, \mathbf{Q} \in \mathbb{C}^{n \times p}$, the complexity of computing $\mathbf{PQ}$ is $\mathcal{O}(m n p)$\cite{boyd2004convex}. Hence, the complexity of computing $\mathbf{Y}_k^{opt}\left(\mu \right)$   and   $\mathbf{\Gamma}$  is  ${\cal O}(KN_t^2D_k)$. Third, similarly,  the complexity of computing $\mathbf{M}$ is ${\cal O}(KD_k^2N_t)$. 
%The complexity of evaluating the Lagrangian multipliers $\left\{\lambda_k, \forall k\right\}$ and $\left\{\mu_l, \forall k\right\}$ can be ignored.
By ignoring the computation complexity in the outer layer, the overall complexity of calculating the beamforming matrices $\mathbf{B}^{(n+1)}$ is  $o_3 = \mathcal{O}\left(I^o_{max}I_{max}\left( KN_{\rm{ t}}^3+KN_{\rm{ t}}^2D_k  +KD_k^2N_{\rm{ t}} \right) \right)$.

The computational complexity of solving Problem (\ref{theta_SDR}) mainly lies in the interior point method \cite{ben2001lectures} and the general expression of which is given by 
\begin{equation} 
	\mathcal{O}\left(\left(\sum_{j=1}^J b_j+2 I\right)^{1 / 2} n(n^2+\underbrace{n \sum_{j=1}^J b_j^2+\sum_{j=1}^J b^3}_{\text {due to LMI }}+\underbrace{ \sum_{i=1}^I a_i^2}_{\text {due to SOC }})\right),\nonumber
\end{equation}
where $n$ is the number of variables, $J$ is the number of linear matrix inequality (LMI) of dimension $b_j$, and $I$ is the number of second-order cone (SOC) of dimension $a_i$  \cite{9180053}. Problem (\ref{theta_SDR}) contains $2$ LMIs with the dimension of $L$ and $1$ LMI with the dimension of $1$. The number of variables is $n = L^2$.  The approximate complexity of solving Problem (\ref{theta_SDR}) is $o_4=\mathcal{O}\left( L^{7.5}\right)$. The
overall algorithm’s computational complexity is $n^{max}\left(o_1 + o_2 + {n_{1}^{max}o_3 + n_{2}^{max}o_4 }\right) $, where $n^{max}$ is the iterations number of the AO algorithm. $n_{1}^{max}$ and $n_{2}^{max}$  represent the number of iterations of optimizing beamforming matrices and phase shift matrix, respectively.
%we employ the standard Gaussian randomization method to get an approximate solution to the problem in (\ref{theta_SDR}) and further get the  phase shift vector through $\mathbf{\Theta}=\operatorname{diag}\{{\boldsymbol{\vartheta}}\}$.
%, for which the details are similar to that in \cite{wu2019intelligent}.
%\begin{algorithm}
%	\caption{MM-based algorithm for solving problem (\ref{otp-theta}).}\label{alg-3}
%	\begin{algorithmic}[1]
%		\STATE {Initialize the iteration number $t= 0$, the maximum number of iterations $t_{max}, $ Given ${\mathbf{U}_k,\mathbf{W}_k },\mathbf{B}_k $. Input the feasible solution $\mathbf{\Theta}^{\left( t\right) }$;}
%		\STATE {\textbf{Repeat}}
%		\STATE {\quad Get $\mathbf{\Theta}^{\left( t+1\right) }$ by solving Problem(\ref{theta_SDR}) with fixed $\mathbf{\Theta}^{\left( t\right) }$, $ t = t+1$;}
%		\STATE {\textbf{Until} $t = t_{max}$;}
%	\end{algorithmic}
%\end{algorithm}

%through Algorithm \ref{alg-3}
\section{SIMULATION RESULTS}\label{simulation}
In this section, we evaluate the performance of the RIS-aided ISAC MU-MIMO systems through simulation. The large-scale path-loss in $\mathrm{dB}$ is given by
\begin{equation}
	\mathrm{PL}=\mathrm{PL}_0-10 \alpha \log _{10}\left(\frac{d}{d_0}\right).
\end{equation}
We set the path-loss exponents of the RIS-related links to $\alpha_{\mathrm{BI}} = \alpha_{\mathrm{IU}} =2.2$. 
%BS-RIS link and of the RIS-CUs links to $\alpha_{\mathrm{BI}} = \alpha_{\mathrm{IU}} =2.2$. 
Considering the presence of numerous obstacles and scatters, we have configured the path-loss exponent for the BS-CUs links as $\alpha_{\mathrm{BU}}=3.75$. $\mathrm{PL}_0=-30 \mathrm{~dB}$  denotes the pathloss at a distance of $1\mathrm{~m}$. For the CUs-related channels, the small-scale fading is assumed to be Rayleigh fading due to extensive scatters. For the direct channel between  BS and RIS, the small-scale fading is assuming to be Rician fading which can be  described as
\begin{equation}
	\mathbf{G}=\sqrt{\frac{K_R}{1+K_R}} \mathbf{G}^{\mathrm{LoS}}+\sqrt{\frac{1}{1+K_R}} \mathbf{G}^{\mathrm{NLoS}},
\end{equation}
where Rician factor  $K_R=3$. 
%where $\mathrm{PL}_0$ is the path-loss at the reference distance $d_0, d$ is the link distance, $\alpha$ is the path-loss exponent. In the simulations, we set $\mathrm{PL}_0=-30 \mathrm{~dB}$ and $d_0=1 \mathrm{~m}$. We set the path-loss exponents of the BS-RIS link and of the RIS-CUs links to $\alpha_{\mathrm{BI}} = \alpha_{\mathrm{IU}} =2.2$, and the pathloss
%exponent of the BS-CUs links are set to $\alpha_{\mathrm{BU}}=3.75$ due to extensive obstacles and scatterers. For the CUs-related channels, the small-scale fading is assumed to be Rayleigh fading due to extensive scatters. Correspondingly, for the direct channel between  BS and RIS, the small-scale fading is assuming to be Rician fading. In specific, the small-scale channel can be modeled as
%\begin{equation}
%	\mathbf{G}=\sqrt{\frac{K_R}{1+K_R}} \mathbf{G}^{\mathrm{LoS}}+\sqrt{\frac{1}{1+K_R}} \mathbf{G}^{\mathrm{NLoS}},
%\end{equation}
%where $K_R$ is the Rician factor, 
${\mathbf{G}}^{\text {LoS }}$ is the deterministic LoS, and ${\mathbf{G}}^{\text {NLoS }}$ is the NLoS Rayleigh component. ${\mathbf{G}}^{\mathrm{LoS}}$ is given by ${\mathbf{G}}^{\mathrm{LoS}}=\mathbf{a}_{D_{\rm{ r}}}\left(\vartheta^{A o A}\right) \mathbf{a}_{D_{\rm{ t}}}^H\left(\vartheta^{A o D}\right)$, where $\mathbf{a}_{D_{\rm{ r}}}\left(\vartheta^{A o A}\right)$ and $\mathbf{a}_{D_{\rm{ t}}}^H\left(\vartheta^{A o D}\right)$ are defined like (\ref{a}). ${D_{\rm{ r}}}$ and ${D_{\rm{ t}}}$ are the number of antennas/elements at the receiver side and transmitter side, respectively. $\vartheta^{D o A}$ is the angle of departure and $\vartheta^{A o A}$ is the angle of arrival. To simplify, we assume that $d / \lambda=1 / 2$. 
%Since the  RIS can be regarded as a monostatic multiple-input and multiple-output (MIMO) Radar, the path-loss $\eta$ in the RIS-target-RIS link can be modelled as
%\begin{equation}
%	\eta=\sqrt{\frac{\lambda^2 S}{(4 \pi)^3 R^4}} .
%\end{equation}
%where $\lambda$ denotes the wavelength of the Radar signal, $S$ denotes the Radar cross section (RCS) of the target, and $R$ denotes the distance between Radar and the target. We set the carrier frequency of the system as $2.7 \mathrm{GHz}$ and the RCS of the target as $100 \mathrm{~m}^2$\cite[Table 2.1 ]{4815550}. 
%Since the target is in a crowded area and pathloss is very large, the parameter of  amplitude of the target $\eta$ is set to $0.1$. 

%\begin{figure*}
%	\begin{minipage}[H]{0.5\linewidth}
%		\centering
%		\includegraphics[scale=0.6,trim=90 260 70 260]{./fig/Opt_B.pdf}
%		\captionsetup{font={small,stretch=1.25}}
%		\caption{Convergence behavior of Algorithm \ref{alg-2}}
%		\label{fig3}
%	\end{minipage}%
%	\begin{minipage}[H]{0.5\linewidth}
%		\centering
%		\includegraphics[scale=0.6,trim=80 260 80 260]{./fig/overall_Conv.pdf}
%		\captionsetup{font={small,stretch=1.25}}
%		\caption{Convergence behavior of  Algorithm \ref{al_overall}}
%		\label{fig4}
%	\end{minipage}
%\end{figure*}
\begin{figure}[]
	\centering
	\setlength{\belowcaptionskip}{6mm}
	\includegraphics[scale=0.8,trim=90 20 70 0]{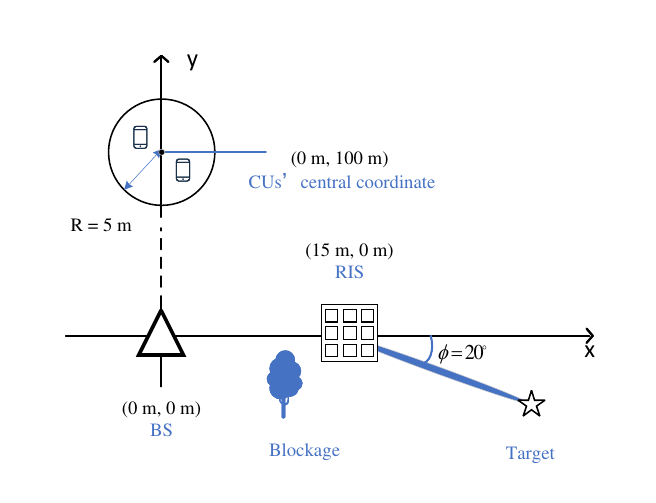}
%	\captionsetup{font={small,stretch=1.25}}
	\caption{\centering{The simulation system scenario.}}\vspace{-0.8cm}
	\label{fig2}
\end{figure}
\begin{figure}
	\centering
	\setlength{\belowcaptionskip}{6mm}
	\includegraphics[scale=0.6,trim=90 260 70 260]{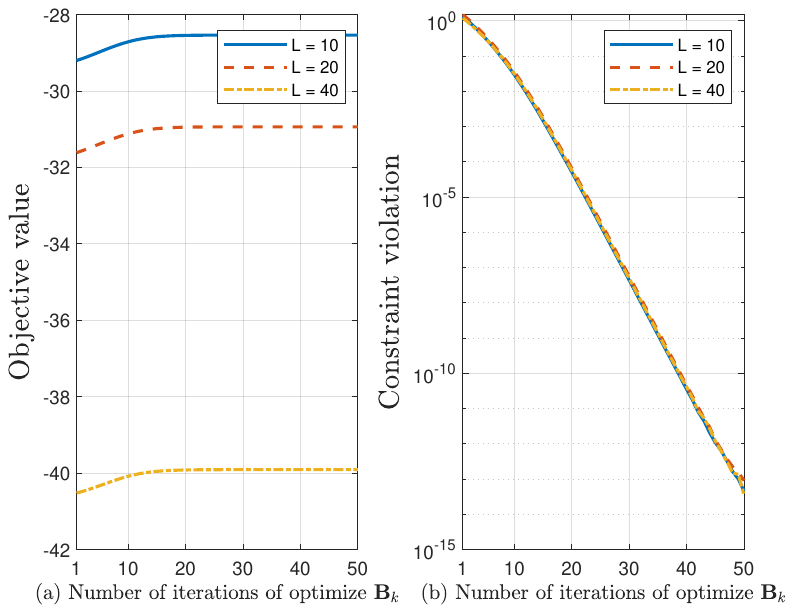}		
%	\captionsetup{font={small,stretch=1.25}}
	\caption{\centering{Convergence behavior of Algorithm \ref{alg-2}.}}\vspace{-0.8cm}
	\label{fig3}
\end{figure}

Unless specified otherwise, we have chosen the following simulation parameters: a channel bandwidth of $10 \mathrm{~MHz}$, a noise power density of $-174 \mathrm{~dBm} / \mathrm{Hz}$, antenna configurations of $N_{\rm{ t}}= N_{\rm{ r}}=4$ for the BS and $M_k=2$ for all CUs.  The number of reflecting  elements is $L=20$,  and the maximum transmit power is $P_{0}=1 \mathrm{~W}$. We assume that the BS and the RIS are located at $(0\mathrm{~m}, 0\mathrm{~m}) $, $ (15\mathrm{~m}, 0\mathrm{~m})  $, respectively. The target is located at a spatial angle of $\phi =-20^{\circ}$  w.r.t RIS, at a distance of $ 5\mathrm{~m} $.
%A target is located at $[15, 5]\mathrm{~m} $ where the spatial angle   $\phi =90^{\circ}$ with channel power gain $|\eta|^{2}=10\mathrm{~dB}$.  
The radar
SNR threshold is set to $\gamma_{0} = 30 \mathrm{~dB}$, CUs are randomly distributed within a circular region centered at $(0\mathrm{~m},100\mathrm{~m})$ and having a radius of $5\mathrm{~m}$.

%Unless otherwise stated, we set the simulation parameters as  follows: Channel bandwidth of $10 \mathrm{~MHz}$, noise power density of $-174 \mathrm{~dBm} / \mathrm{Hz}$, the numbers of antennas for the dual-function BS and the CUs are $N_t= N_r=4$, and $M_k=2, \forall k$, respectively. Number of reflecting  elements  $L=20$, the maximum transmit power  $P_{0}=1 \mathrm{~W},$ Rician factor  $K_R=3$. We assume that the BS and the RIS  are located at $[0, 0]\mathrm{~m} $, $ [15, 0] \mathrm{~m} $, respectively. 
%A target is located at a spatial angle of $\phi =-20^{\circ}$  relative to RIS, at a distance of $ 5\mathrm{~m} $.
%%A target is located at $[15, 5]\mathrm{~m} $ where the spatial angle   $\phi =90^{\circ}$ with channel power gain $|\eta|^{2}=10\mathrm{~dB}$.  
%The radar
%SNR threshold is set to $\gamma_{0} = 30 \mathrm{~dB}$, the users are in a circle with the central coordinate of $[0,100] \mathrm{~m}$ and the radius of $5\mathrm{~m}$. 

\begin{figure}
	\centering
	\setlength{\belowcaptionskip}{6mm}
	\includegraphics[scale=0.6,trim=90 260 70 260]{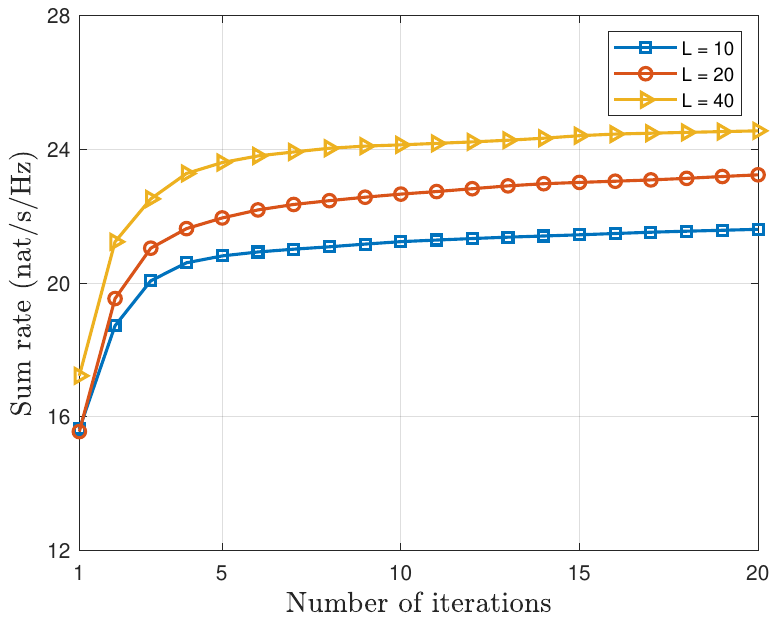}		
%	\captionsetup{font={small,stretch=1.25}}
	\caption{\centering{Convergence behavior of  Algorithm \ref{al_overall}.}}\vspace{-0.8cm}
	\label{fig4}
\end{figure}
We consider the following benchmark schemes to  evaluate the  RIS-aided DFRC system.  We denote ‘Proposed’ as  our proposed approach described in Algorithm \ref{al_overall}. In the  Random RIS scenario (denoted by ‘Random’), the phase shift matrix that satisfies the constraints are randomly generated during initialization. Subsequently, the optimization process focuses exclusively on optimizing the beamforming matrix. Besides, ‘Com-only’ represents that the system only performs the communication function without considering the radar SNR constraint.

%We consider the following benchmark scenario to fully evaluate the  RIS-aided ISAC system.  We denote ‘Proposed’ as  our proposed approach described in Algorithm \ref{al_overall}. The scenario Random RIS, denoted by ‘Random’ represents that the RIS phase shifts are generated randomly following a uniform distribution over $[0, 2\pi)$ and  constraint (\ref{SNR}) are met. Specifically, beamforming matrices and a phase shift matrix that satisfies the constraints are randomly generated during initialization, and then only the beamforming matrices are optimized later. Besides, ‘Com-only’ represents that the system only performs the communication function without radar SNR constraint.

\subsection{Convergence behavior of the proposed algorithm}

First, we evaluate the convergence of the used penalty-based algorithm on optimizing the  transmit beamforming matrix $\mathbf{B}_k$. Penalty-based algorithm may require careful adjustment of the penalty coefficient $\rho$ and the updated step size $c$ to ensure convergence. The selection of inappropriate  parameters may lead to the non-convergence of the optimization problem and the computational workload for solving the problem may substantially rise,  resulting in extended computation time, particularly for complex problems. To evaluate the trade-off  between convergence performance and computational complexity of Algorithm \ref{alg-2}, we plot Fig. \ref{fig3} with carefully selected $\rho = 100$ and  $c = 0.7$. Fig. \ref{fig3}(a) illustrates the monotonic increase in the objective value (\ref{obj_value}) with varying numbers of RIS reflecting elements, specifically $L = 10$, $L = 20$, and $L = 40$, as the number of outer layer iterations grows. With the appropriate parameters $\rho$ and $c$, it is clear that the objective function reaches convergence after approximately $15$ iterations.
%shows the constraint violation and convergence of Algorithm \ref{alg-2} by solving Problem (\ref{ADMM_X_k})  with different numbers of RIS reflecting elements, i.e., $L = 10$, $L = 20$, and $L = 40$. 
%Fig. \ref{fig3} shows the constraint violation and convergence of Algorithm \ref{alg-2} by solving Problem (\ref{ADMM_X_k})  with different numbers of RIS reflecting elements, i.e., $L = 10$, $L = 20$, and $L = 40$. 
The penalty terms $ \sum_{k=1}^K\Vert {\mathbf{X}_k} - \mathbf{B}_{k}\Vert_{F}^{2}$  and $\sum_{k=1}^K\Vert {\mathbf{Y}_k} -  \mathbf{V}\mathbf{B}_{k}\Vert_{F}^{2}$ decrease rapidly with the number of outer layer iterations as  shown in Fig. \ref{fig3}(b). As $\rho$ becomes very small,  the penalty terms achieve $10^{-13}$ after approximately 50 iterations, indicating  the equality constraint (\ref{auxiliary}) in Problem (\ref{auxiliary_X_Y}) is eventually satisfied. Thus, Algorithm \ref{alg-2} is guaranteed to converge finally.

%As shown in Fig. \ref{fig3}(b), it's evident that the constraint violation decreases rapidly and achieves the predefined accuracy of $10^{-13}$ after approximately 50 iterations for $L = 40.$ This indicates that $\sum_{k=1}^K\Vert {\mathbf{X}_k} - \mathbf{B}_{k}\Vert^{2}$ and $\sum_{k=1}^K\Vert {\mathbf{Y}_k} - \mathbf{V}\mathbf{B}_{k}\Vert^{2}$ in (\ref{Problem_with_auxi_in_obj}) are driven to approach zero.
%From Fig. \ref{fig3}(b), it is observed that constraint violation decreases fast and reaches the predefined accuracy $10^{-13}$ after about $50$ iterations for $L = 40$, which indicates that $ \sum_{k=1}^K\Vert {\mathbf{X}_k} - \mathbf{B}_{k}\Vert^{2}$  and $\sum_{k=1}^K\Vert {\mathbf{Y}_k} -  \mathbf{V}\mathbf{B}_{k}\Vert^{2}$ in (\ref{Problem_with_auxi_in_obj}) are forced to approach zero. 
%As such, the equality constraint (\ref{auxiliary}) in Problem (\ref{auxiliary_X_Y}) is eventually satisfied. 
%By adjusting the appropriate parameters $\rho$ and $c$, in Fig. \ref{fig3}(a), we can observe that the objective value (\ref{obj_value}) is  monotonically increasing with the number of outer layer iterations.
%As $\rho$ becomes very small, the constraint violation is forced to approach the predefined accuracy. 

In Fig. \ref{fig4}, we investigate the convergence characteristics of Algorithm \ref{al_overall} versus the number of RIS reflecting elements.
%In Fig. \ref{fig4}, we study the convergence behavior of the   overall  Algorithm \ref{al_overall} for different number of RIS reflecting  elements $L$. 
It is observed from Fig. \ref{fig4} that the sum rate achieved for various $L$ values increases monotonically with the number of iterations. Additionally, the algorithm converges rapidly and in general 20 iterations are sufficient for the  algorithm to achieve a large portion of the converged sum rate. 

\subsection{Impact of the number of reflective elements on the sum rate}
\begin{figure}
	\centering
		\setlength{\belowcaptionskip}{6mm}
	\includegraphics[scale=0.6,trim=90 260 70 260]{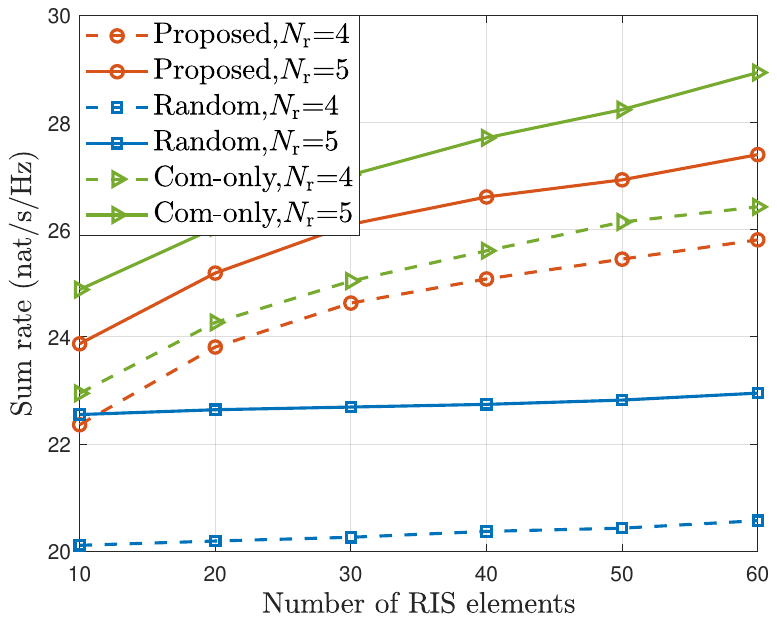}		
%	\captionsetup{font={small,stretch=1.25}}
	\caption{\centering{Sum rate versus the number of RIS elements $ L$ .}}\vspace{-0.8cm}
	\label{fig5}
\end{figure}

In Fig. \ref{fig5}, we demonstrate the relationship between the achievable sum rate and the number of reflecting elements $L$. As increasing of the number of RIS reflecting  elements, a larger sum rate is achieved because of a larger passive beamforming gain.  As the number of reflecting elements increases, the performance gap between the the ‘Proposed’ scheme and the ‘Random’ scheme increases. This indicates that it is essential to optimize the phase shift of the RIS especially with a large number of the reflecting elements. Moreover,  
%相比于COM系统，DFRC系统在对通信影响较小的情况下，增加了感知功能，同时，两者之间的gap也体现了通信和感知之间的权衡
the DFRC system adds sensing capabilities while having minimal impact on communication than the ‘com-only’ system. At the same time, the gap between the two also reflects the trade-off between communication and radar sensing performance.

\begin{figure}
	\centering
	\setlength{\belowcaptionskip}{6mm}
	\includegraphics[scale=0.6,trim=90 260 70 260]{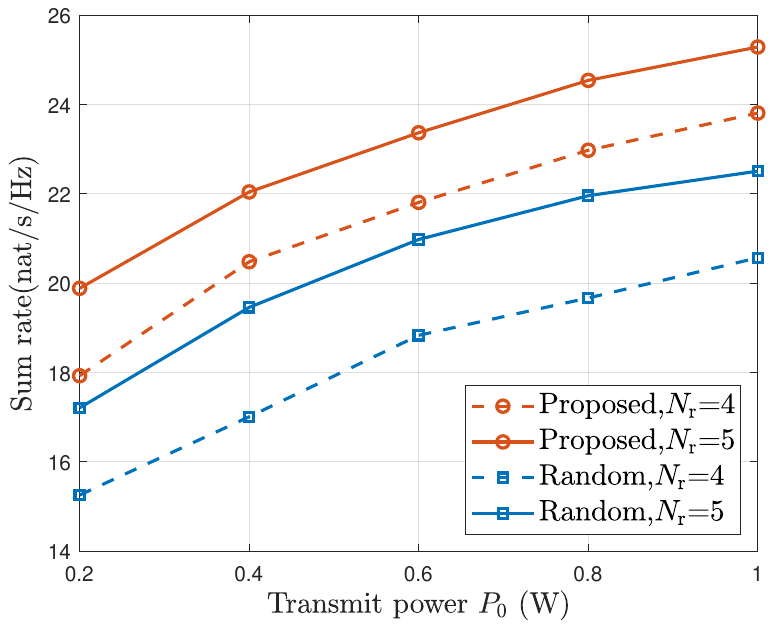}		
%	\captionsetup{font={small,stretch=1.25}}
	\caption{\centering{Sum rate versus $P_0$.}}\vspace{-0.8cm}
	\label{fig6}
\end{figure}

\subsection{Impact of the BS transmit power on the sum rate}
In this section, we set the number of RIS reflecting  elements to $L =20 $ and evaluate the relationship between the sum rate and the maximum transmit power of the dual-function BS. In Fig. \ref{fig6}, we observe that the proposed approach achieves a remarkable performance improvement compared to the ‘Random’ scheme. In addition, the scenarios with $N_{\rm{ r}} = N_{\rm{ t}}=5$ achieve better performance than their counterparts with $N_{\rm{ r}} = N_{\rm{ t}}=4$ thanks to higher beamforming gain.
%\begin{figure}[H]
%	\centering
%	\includegraphics[scale=0.6]{./fig/Var_P02-1.pdf}
%	\caption{\centering{Sum-rate versus $P_0$ .}}\vspace{-0.8cm}
%	\label{fig6}
%\end{figure}
\subsection{Impact of the BS-CUs path-loss exponent on the  sum rate}

\begin{figure}
	\centering
	\setlength{\belowcaptionskip}{6mm}
	\includegraphics[scale=0.6,trim=90 260 70 260]{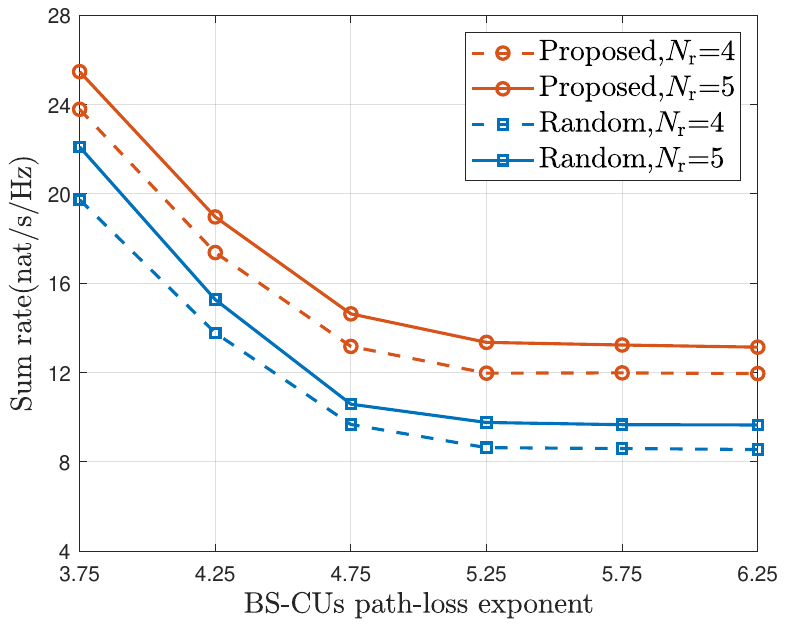}	
%	\captionsetup{font={small,stretch=1.25}}
	\caption{\centering{Achievable sum rate versus $\alpha_{\mathrm{BU}}$.}}\vspace{-0.8cm}
	\label{fig8}
\end{figure}

In the preceding examples, we have configured the path-loss exponents for the BS-CUs links as $\alpha_{\mathrm{BU}}=3.75$. However, in some practical scenarios, especially in an urban environment, surrounding buildings are likely to hinder the BS-CUs path. To this end,  Fig. \ref{fig8} is plotted to evaluate the relationship between the sum rate and the BS-CUs path-loss exponents. The sum rate achieved by all scenarios  decreases as $\alpha_{\mathrm{BU}}$ increases, eventually it reaches a plateau with no significant further decrease. This is due to the fact that with the increase of  $\alpha_{\mathrm{BU}}$, the signal received at the BS is weaker. However, when $\alpha_{\mathrm{BU}}$ is larger than 5.25, the BS-CUs link cannot operate normally due to severe path loss. At this time, the reflection link provided by RIS can ensure that the sum rate is not less than $8$ $\mathrm{nat/s/Hz}$.

%当α大于5.25时，BS-CUs 链路因为损耗太大而无法正常工作，此时RIS提供的反射链路可以保证和速率不低于8bit/s/Hz
%This section evaluates the relationship between the sum rate and the BS-CUs path-loss exponent. In the above examples, the path-loss exponents of the BS-CUs links are set as $\alpha_{\mathrm{BU}}=3.75$, however, in some practical scenarios, especially in an urban environment, surrounding buildings are likely to hinder BS-CUs path. Hence, it is intriguing to investigate the performance gain that can be achieved by our proposed algorithms when the BS-CUs link experience rich scattering fading with a higher value of $\alpha_{\mathrm{BU}}$. To this end, we plot Fig. \ref{fig8} to show the impact of the BS-CUs path-loss exponent on  sum rate. The sum rate achieved by all scenarios  decreases upon increasing $\alpha_{\mathrm{BU}}$, and eventually leveled off. This is because upon increasing $\alpha_{\mathrm{BU}}$, the signal attenuation associated with the BS-CUs links becomes larger, and the signal received from the BS is weaker, hence more negligible. 

%However, no matter how large parameter $\alpha_{\mathrm{BU}}$ is, significant performance gains can be achieved by our proposed algorithms over the Random scenario, which shows the significance of RIS in improving the communication QoS, especially in congested urban environments with high path loss.

\subsection{Impact of the number of quantization bits on the sum rate}
In our simulation, we assume that the phase shifts of RIS elements are continuous. However, in practical systems, continuous phase shifts can bring high costs and may even be impossible to achieve. Therefore, discretizing the continuous phase shift of RIS is a more economical method. Specifically, we define $B$ as the number of  quantization bits and  the corresponding set of discrete phase shifts is 
\begin{equation}\mathcal{A} = \left\lbrace \frac{2 \pi m}{2^{B}} \right\rbrace ,m \in [0,2^{B}-1],m\in \mathbb{N}.
\end{equation}

We define $\left\lbrace \phi^{\star}(n)\right\rbrace $ as the optimal continuous phase obtained by the proposed algorithm, and then the discrete phase shifts can be obtained as follows:
\begin{equation}
	\hat{\phi}(n) = {\mathop{\operatorname{argmin}}\limits_{\left\lbrace \phi\in\mathcal{A}\right\rbrace }}\left\lbrace \left|\phi- \phi^{\star}(n)\right| \right\rbrace .
\end{equation}

\begin{figure}
	\centering
	\setlength{\belowcaptionskip}{6mm}
	\includegraphics[scale=0.6,trim=80 260 70 260]{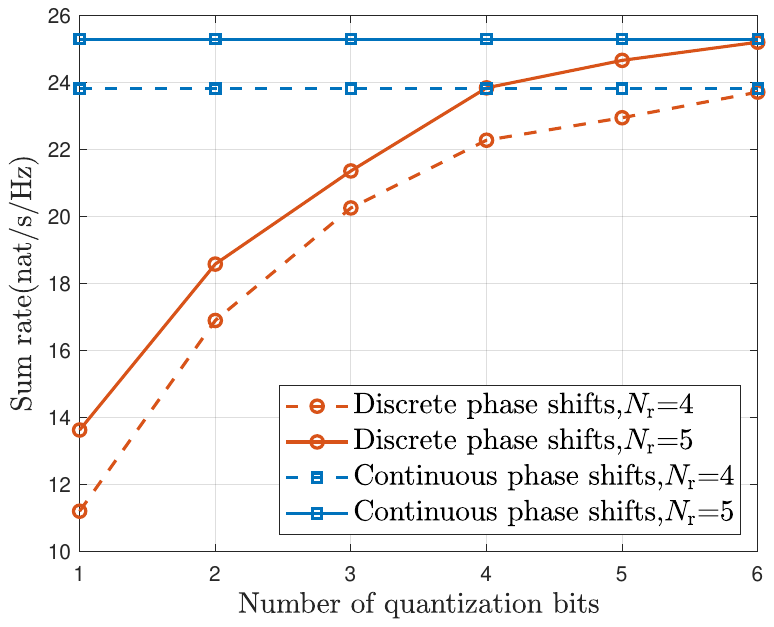}		
%	\captionsetup{font={small,stretch=1.25}}
	\caption{\centering{Sum rate versus the number of quantization bits $B$.}}\vspace{-0.8cm}
	\label{fig9}
\end{figure}

As shown in Fig. \ref{fig9},  we present the effect of the number of quantization bits on the sum rate. We can see that when $B$ is small, the sum rate increases rapidly with the increase of $B$, while when $B$ is large,  the increase of the curve tends to be gradually saturated and close to the sum rate under continuous conditions. The figure shows that six quantization bits can achieve a large portion of that of  continuous  phase shifts. This shows that the proposed algorithm can effectively reduce hardware cost and power consumption in practical systems.

\subsection{Transmit Beampattern}
\begin{figure}[]
	\centering
	\setlength{\belowcaptionskip}{6mm}
	\includegraphics[scale=0.6,trim=80 260 70 260]{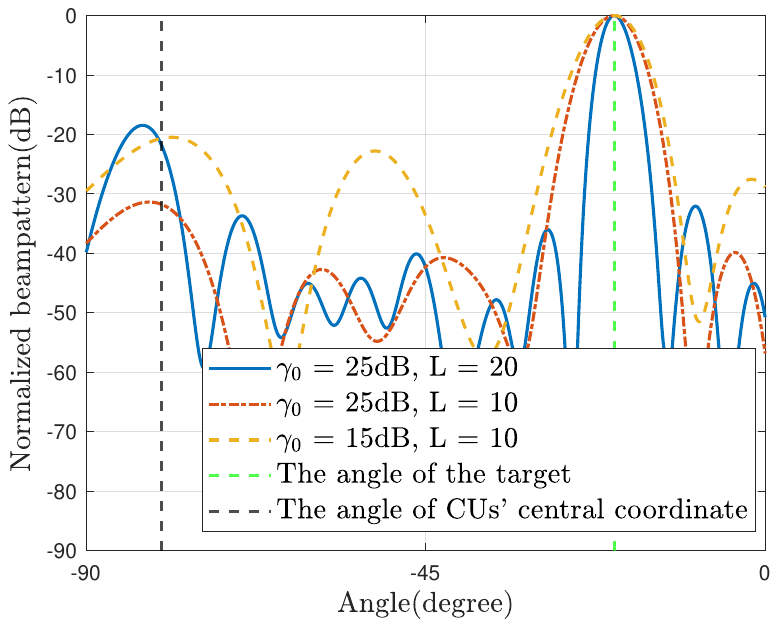}
%	\captionsetup{font={small,stretch=1.25}}
	\caption{\centering{Transmit Beampattern.}}\vspace{-0.8cm}
	\label{fig7}
\end{figure}

Fig. \ref{fig7} presents a comparison of the normalized sensing beampattern for different numbers of reflecting elements of the RIS and diverse radar performance requirements. The beampattern gain from the  RIS towards angle $\theta$ is defined as    
\begin{equation}
		P_{}(\theta)  =\left|\mathbf{a}(\theta)^{\rm{ H}} {\mathbf{\Theta}}^{\rm{H}} {\bf{G}} 	{{\bf{x}}} \right|^2.
\end{equation}
Firstly, we compared the beampatterns with $\gamma_0=15\mathrm{~dB}$ and $\gamma_0=25\mathrm{~dB}$ when $L = 10$, and it can be seen that the latter has lower sidelobe.  Meanwhile, if we increase the number of elements in RIS by $L=20$, we can observe that the mainlobe becomes narrower, which  further illustrates the benefit of increasing the number of reflecting elements of RIS on sensing performance. Under the same sensing requirements, more power is {illuminated in the direction of CUs} when  the number of the reflecting element is increased to $L = 20$.

%Compared with the case of $\gamma_0=15\mathrm{~dB}$, more energy in the case of $\gamma_0=25\mathrm{~dB}$ is  used to meet the sensing requirements, and the corresponding communication beam is not ideal. 
%Under the same sensing requirements, increasing the number of reflecting elements of RIS to $L = 20$, it can be observed
%that more power is illuminated in the direction of CUs, due to the sensing constraints can be more easily satisfied.

\section{CONCLUSION}\label{Conclusion}

In this paper, we studied an RIS-aided MU-MIMO DFRC system. We aimed to maximize
the achievable sum rate of the CUs by jointly optimizing the BS beamforming matrix and the passive reflecting coefficients of the RIS while satisfying radar SNR constraint, the transmit power constraint of BS  and the unit modulus property of the reflecting coefficients of RIS. An AO approach was employed to decouple the optimization variables and split this intractable problem into two subproblems. These subproblems were addressed individually using a penalty-based algorithm and an MM-based algorithm, respectively. Simulation results demonstrated the significant advantages of deploying RIS in ISAC systems. In the future, we will extend our work to active RIS-aided ISAC systems due to the superiority it indicated in\cite{10319318,zhi2022active}.

\bibliographystyle{IEEEtran}{}
\bibliography{first_2022_11}
% \end{multicols}
\end{document}